%% file: main.tex
\documentclass{scrartcl}

\usepackage[natbibapa]{apacite}
\usepackage[utf8]{inputenc}
\usepackage[T1]{fontenc}
\usepackage{amsmath}
\usepackage{amsfonts}
\usepackage{amssymb}
\usepackage{lmodern}
\usepackage{color}
\usepackage{graphicx}
\usepackage[lmargin=1cm,rmargin=1.5cm]{geometry}
\usepackage{lscape}
\usepackage{multicol}
\usepackage{subcaption}
\usepackage{empheq}
\usepackage{amsmath}
\usepackage{gensymb}
\usepackage{hyperref}

\usepackage{setspace}
\usepackage{lineno}

\RequirePackage{silence}
\usepackage{array}
\newcolumntype{x}[1]{>{\centering\let\newline\\\arraybackslash\hspace{0pt}}p{#1}}

\title{\LARGE\textbf
Bayesian Dynamical Modeling of Fixational Eye Movements}
\author{Lisa Schwetlick$^{1}$, Sebastian Reich$^{2,3}$, \& Ralf Engbert$^{1,3}$ \\ $^1$Department of Psychology,  \\$^2$Institute of Mathematics \\$^3$Research Focus Cognitive Sciences\\ 
University of Potsdam, Germany}

\begin{document}
\maketitle

\abstract{
Humans constantly move their eyes, even during visual fixations, where miniature (or fixational) eye movements occur involuntarily. Fixational eye movements comprise slow components (physiological drift and tremor) and fast components (microsaccades). The complex dynamics of physiological drift can be modeled qualitatively as a statistically self-avoiding random walk \citep[SAW model,][]{Engbert2011}. In this study, we implement a data assimilation approach for the SAW model to explain statistics of fixational eye movements and microsaccades in experimental data obtained from high-resolution eye-tracking. We discuss and analyze the likelihood function for the SAW model, which allows us to apply Bayesian parameter estimation at the level of individual human observers. Based on model fitting, we find a relationship between the activation predicted by the SAW model and the occurrence of microsaccades. The model's latent activation relative to microsaccade onsets and offsets using experimental data lends support to the existence of a triggering mechanism for microsaccades. Our findings suggest that the SAW model can capture individual differences and serve as a tool for exploring the relationship between physiological drift and microsaccades as the two most essential components of fixational eye movements. Our results contribute to understanding individual variability in microsaccade behaviors and the role of fixational eye movements in visual information processing.}

\vspace{1cm}

\newpage


The human eye is never truly at rest. At a macro-level the eyes move in a sequence of fixations and saccades \citep{Schwetlick2020a}, moving different aspects of the visual world into the receptor-dense center of the visual field (i.e., the fovea). Despite the misleading term, the eyes are far from stationary during fixations \citep{Ditchburn1959, Kowler2011, MartinezConde2004, Rucci2015, Poletti2023}. Microscopic or fixational movements constantly shift the visual input over the receptors in the retina. Three components of fixational eye movement behavior are distinguished: a slow random motion called physiological drift, a high-frequency tremor, and high-velocity microsaccades \citep{Alexander2019,Ciuffreda1995}. The comparatively small amplitude of tremor movement can not be resolved in current video-based eye tracking and is therefore neglected in the following. 

The function of fixational eye movements, their underlying mechanisms, as well as the consequences for processing in the visual system have yet to be fully understood. While fixational eye movements are found ubiquitously across species and in all primates \citep{Ko2016}, The generation of fixational eye movements varies between individuals and is highly characteristic 
\citep{Engbert2003,Engbert2004,Cherici2012, Poynter2013}. 
To model the individual spatial statistics of physiological drift and its relationship to microsaccades, we briefly discuss the two movement components.

During physiological drift, the eyes move in a pattern that resembles Brownian motion \citep{Engbert2004,Engbert2006a,Burak2010,Pitkow2007}, i.e., eye position meanders stochastically with increasing variance over time. However, a more detailed analysis indicates that fixational eye movements represent an interesting example of fractional Brownian motion \citep{Mandelbrot1968,Metzler2000}. A corresponding analysis \citep{Collins1995} can be carried out by computing the mean square displacement (MSD) at different time lags \citep{Engbert2004,Herrmann2017}. In Brownian motion, the MSD increases linearly with time lag. In fixational eye movements, however, a superdiffusive tendency is found over short time scales ($\lesssim 50$~ms), which is also referred to as persistence. Over longer time scales ($\gtrsim 100$~ms) physiological drift is found to be anti-persistent. We interpret such behavior by assuming that fixational eye movements maximize movement over short time scales to counteract retinal fatigue while reducing variance over a long time scale to maintain visual fixation at an intended region of interest or object. \citep{Engbert2004}. 

Microsaccades share their kinematic properties with their larger counterparts, such as acceleration profile,  main sequence (i.e., the linear relationship between log amplitude and log peak velocity; see \citet{Bahill1975}), and are generated by the same neural circuits in the brainstem \citep{Hafed2009,Hafed2021}. Microsaccades are distinguished mainly by their smaller amplitude, usually thresholded at $<1\degree$ \citep{MartinezConde2004,Poletti2023}. Microsaccades occur at a rate which is highly variable between subjects \citep{Engbert2003,MartinezConde2004}. Attentional mechanisms also modulate microsaccades in both their rate and orientation \cite{Hafed2002,Engbert2003,YuvalGreenberg2014}. The rate variation is induced by display changes and modulated by ongoing cognition: First, a reduced microsaccade rate follows target onset, and then an increased rate \citep{Engbert2003} is observed. Microsaccade orientations are modulated by  covert attention, e.g., during spatial cueing \citep{Hafed2002, Engbert2003}.  The general patterns of interactions of microsaccade rates and orientations is more complex \citep{Engbert2006a}, however, a computational models has been proposed \citep{Engbert2012}.

Early accounts of fixational eye movements conceptualized them as the result of random firing from oculomotor units \citep{Eizenman1985}, or else as a nuisance component that causes blurring, if not corrected by the visual system \citep{Packer1992,Burak2010}. More recent evidence points to fixational eye movement being a necessary and useful component of visual exploration in counteracting receptor adaptation \citep{Rucci2015,Engbert2006,Intoy2020,Poletti2023}. First, fixational eye movements prevent visual fading \citep{Ditchburn1959, MartinezConde2006} caused by neural adaptation \citep{Coppola1996, MartinezConde2004, MartinezConde2006}. Although fading prevention may be achieved by drift alone, microsaccades are much more effective at restoring vision after fading has set in \citep{McCamy2014, McCamy2012}. Second, both drift \citep{Rucci2015, Boi2017} and microsaccades \citep{Poletti2013} have been found to facilitate high acuity pattern vision \citep{Intoy2020}. Specifically, the performance of an edge detection model can be improved by the addition of a movement component \citep{Schmittwilken2022}. In another study, \citet{Anderson2020} use a Bayesian model of neurons during early visual processing that simultaneously estimates eye motion and object shape. This study also reports that drift motions benefit high-acuity vision, mainly by averaging over the inhomogeneities in the retinal receptors and receptor density. Finally, microsaccades and drift have been found to be both corrective (i.e., moving the eyes back to the intended fixation position) and exploratory or error-producing (i.e., moving new details into the center of the visual field) \citep{Engbert2004}. Microsaccades are typically preceded by a reduction in drift \citep{Engbert2006, Sinn2016}. Thus, both microsaccades and drift are functionally related. The current study's research goal is to analyze the relationship between slow fixational eye movement components and microsaccades based on quantitative modeling while taking into account individual differences observed in experimental data.

The SAW model integrates several of the above properties of fixational eye movement and is biologically plausible 
\citep{Engbert2011,Engbert2012,Herrmann2017}. The model describes physiological drift  by assuming a \textbf{S}elf-\textbf{A}voiding (random) \textbf{W}alk (SAW) confined in a movement potential that limits the movements to reproduce visual fixation. A random walk on a lattice represents the trajectory of the eye. As the random walk traverses the lattice, visited locations are activated (see Figure \ref{fig:modelspec}). The activation represents the memory process that keeps track of the recently visited locations \citep{Freund1992}. This generative model successfully reproduces both the persistent and anti-persistent statistical properties of ocular drift \cite{Engbert2011}. An extension of the model \citep{Herrmann2017} implements neurophysiological delays, thereby matching the characteristic oscillations found in the displacement autocorrelation. Within the SAW model framework, \cite{Engbert2011} proposed a mechanism for generating microsaccades based on the activation in the SAW model. However, the model has been a qualitative account, as it did not attempt to reproduce experimental data from human observers quantitatively.

We study the SAW model in a likelihood-based framework in the present work to enable Bayesian parameter inference from fixational eye-movement data. We estimate model parameters for individual observers and find that the estimated parameters represent individual spatial statistics of physiological drift. Based on the quantitative agreement between simulated and experimental drift movements, we explore the relation to microsaccades. We use the model's latent activation to investigate potential mechanisms for triggering microsaccades. Since fixational eye movements strongly impact the spatiotemporal input that the visual system processes,  reproducing these movements from a computer-implemented mathematical model is essential for a better understanding of visual functioning.

\section*{Results}
In the first step, we define the SAW model and describe the computation of its likelihood function. In the second step, we use the likelihood computation to estimate parameters for individual human participants. Finally, we use the model to generate data and conduct posterior predictive checks and exploratory analyses concerning the relationship between drift and microsaccades.

\subsection*{The model}
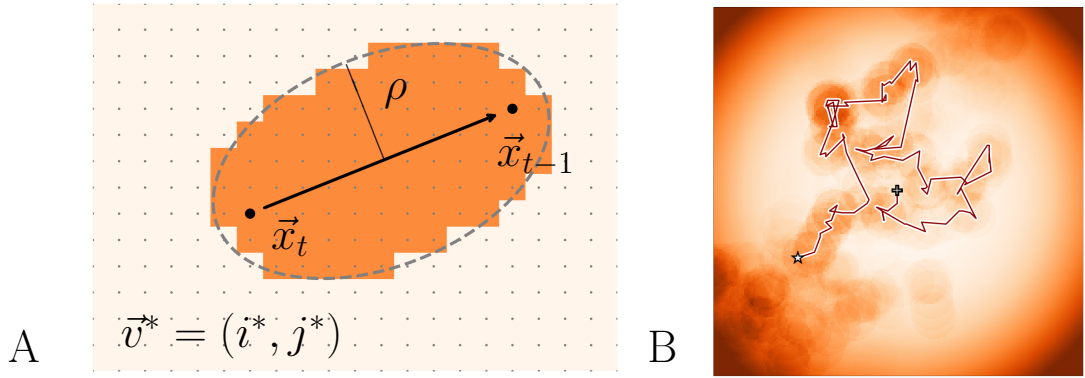
\begin{figure}
\unitlength1mm
\begin{picture}(150,60)
 \put(5, -12){\includegraphics[width=90mm]{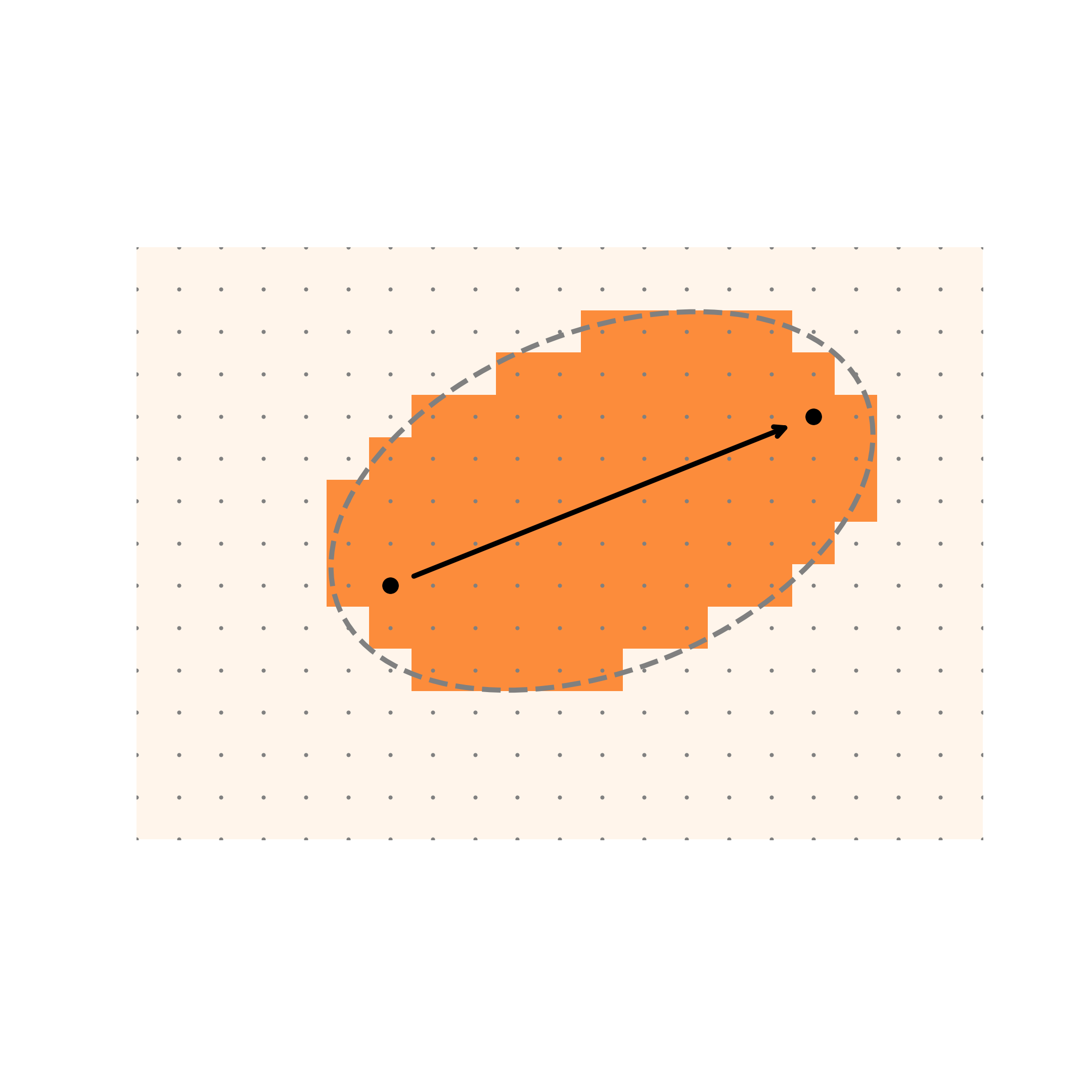}}
 \put(40,25){\LARGE{$\textcolor{black}{\vec{x}_{t}}$}} 
 \put(70,36){\LARGE{$\textcolor{black}{\vec{x}_{t-1}}$}} 
 \put(55,45){\LARGE{$\textcolor{black}{\rho}$}} 
 \put(20,12){\LARGE{$\textcolor{black}{\vec{v}^{*}=(i^{*},j^{*})}$}} 
 \put(5,10){\LARGE{A}} 
 \put(55,37){\line(-2,5){5}}
 
 \put(90, 0){\includegraphics[width=65mm]{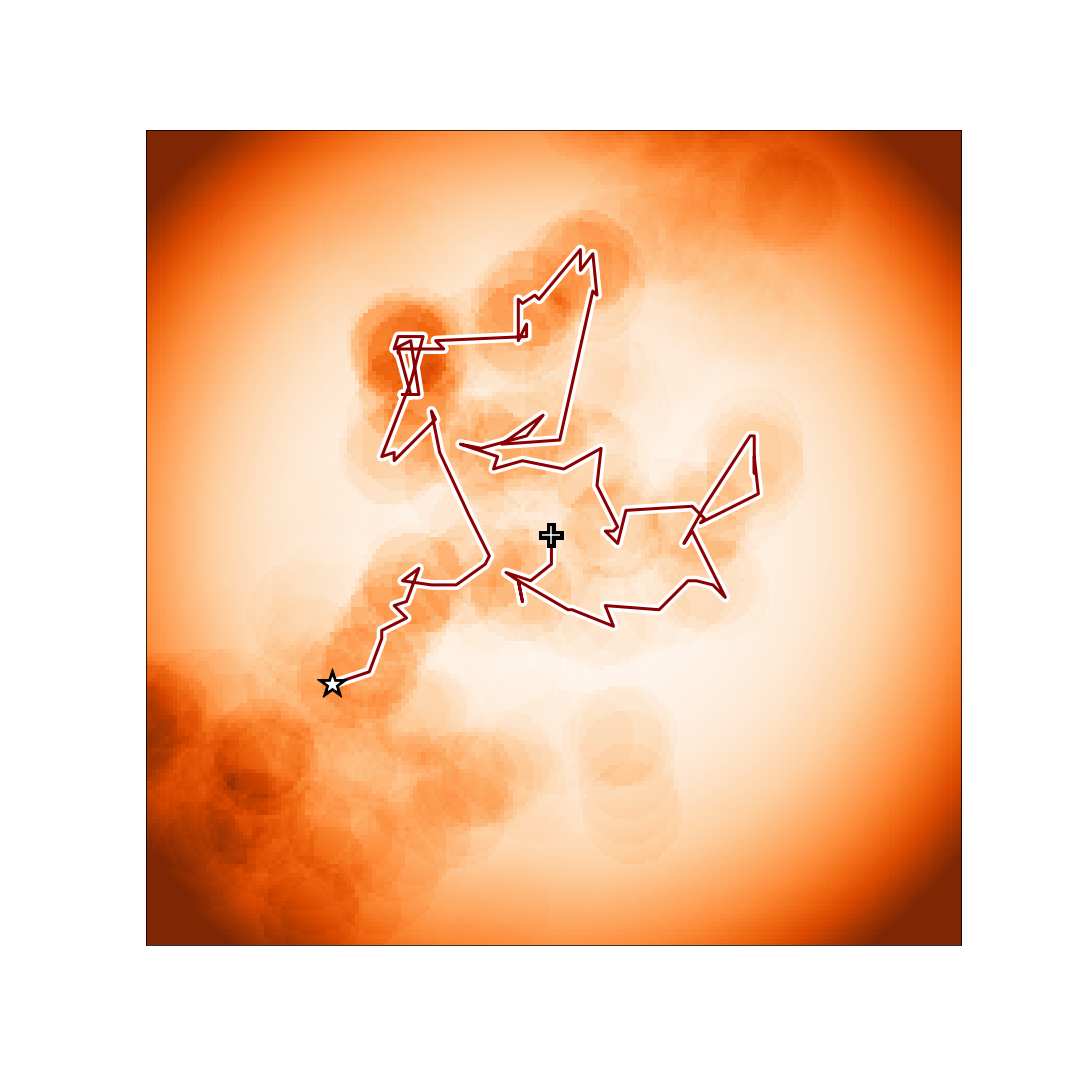}}
 \put(90,10){\LARGE{B}} 
\end{picture}
\caption{\label{fig:modelspec}
    \textbf{A} Activation of lattice points within distance $\rho$ from the eye's trajectory. \textbf{B} Simulated fixational eye movement trajectories, illustrating the SAW model. Starting from the position marked by the + and ending at the position marked by the star, the model generates activation along the trajectory. The movement is constrained by an activation potential centered around the fixation position. The activation profile is initialized by simulating some movement causing activation to be visible at locations that have not been visited. Activation decays gradually over time.
}
\end{figure}

As a theoretical starting point, the SAW model generates a random walk which is statistically self-avoiding \citep{Freund1992}. The self-avoiding walk is implemented on an $L\times L$ lattice where nodes are given by $(i, j)$ with $i,j=1,2,3,...,L$. Each node carries some activation $a_t(i,j)$ at time $t$, which can be interpreted as neural firing rates. Initial activation values are set to $a_{t=0}(i,j) = 10^{-1}$ for all $(i,j)$. At time $t=1,2,3,\dots$, first, the current activation of each node $(i,j)$ across the field decays, according to
\begin{equation}
    a_{t+1}(i,j) = \epsilon \cdot a_{t}(i,j) \;,
\end{equation}
where $\epsilon= 1 - (10^{\gamma})$, representing the speed of the process memory decay. Second, activation is added to the nodes along the walker's trajectory, i.e.,
\begin{equation}
\label{eq:diffeq}
    a_{t+1}(i^\star,j^\star) = a_t(i^\star,j^\star) + 1 \;.
\end{equation}
Next, we implement a rule for activating lattice positions $(i^\star,j^\star)$ along the trajectory. We define an ellipse which is drawn such that the positions at times $t$ and $t+1$ are the foci of the ellipse. The parameter $\rho$ represents the size of the minor axis of the ellipse; the numerical value is set to $12$ units (see Fig.~\ref{fig:modelspec}). The lattice positions $\vec{v}^\star = (i^\star,j^\star)$ are defined as all lattice sites within the ellipse, which are activated according to Eq.~(\ref{eq:diffeq}).

The discretized map $\{a_t(i,j)\}$ of neural activations can be interpreted biologically, since grid cells have been found in entorhinal cortex which keep track of previously visited locations \citep{Killian2012}. 

In principle, the self-avoiding walk can produce persistent motion if parameters are selected appropriately. However, during visual fixation, human observers are able to keep the eyes at an intended target. Therefore, the model implements a movement potential, which confines the random walk and represents a mechanism of fixation control. As a consequence, the model can produce anti-persistent motion on the longer time scale. Thus, the self-avoiding motion in a potential could maintain fixation at the intended location despite the necessity of refreshing the retinal input. The (time-independent) confining potential $u$ is centered in the lattice and takes the form
\begin{equation}
    u(i,j) = \lambda \frac{\left( \sqrt{(i-\frac{L}{2})^2+(j-\frac{L}{2})^2}\right)^\nu}{L^\nu}  \;.
\end{equation}
Within this potential, motion is controlled by the sum of the self-generated activation $a_t(i,j)$ and the potential $u(i,j)$, i.e.,
\begin{equation}
    q_t(i,j)=a_t(i,j)+u(i,j)
\end{equation}

In order to chose the next step of the walker, we compute the probability of the next eye position as $\pi_n$, consisting of $q_t(ij)$, self generated activation and potential, and a time-independent stepping distribution, which controls the size of the movements, i.e.,
\begin{equation}
\label{eq:stepping}
    \pi_t(i,j) = {q_t(i,j)^{-\eta}}\,
    \exp\left(-\left[ \left(\frac{i}{r_i}\right)^\phi + \left(\frac{j}{r_j}\right)^\phi \right]\right) \;.
\end{equation}
From $\pi_t(i,j)$ we select the eye position at time $t+1$ using a linear selection algorithm.

As a result, in each time step follows the sequence of, first, relaxation of the current activation; second, selection of the following eye position under consideration of a stepping distribution; and third, an increase of the activation values along the current trajectory. Several free parameters control our model's behavior. In this study, we selected a subset of parameters for estimation, namely $\gamma$, the speed of the relaxation, the size of the stepping distribution $r_i$ and $r_j$, the slope of the stepping distribution $\phi$ and the slope of the potential $\lambda$ (i.e., $\theta = [\gamma, r_i, r_j, \phi, \lambda]$). We set $\rho = 12$, $\nu = 3$, and $\eta = 1$ to constrain the model and obtain numerically stable behavior during parameter estimation. The parameters selected for estimation are of primary interest, as their values are interpretable quantities that may give insight into the biological plausibility of the model. 

\subsection*{Likelihood function: sequential computation}
The observation that makes the model compatible with a likelihood-based approach is the fact that the likelihood $L$ at each time for each location on the lattice is given by the selection map $\pi$. In other words given the time-ordered data $X_t=\{x_1,x_2\dots,x_t\}$, we can calculate the probability of observing the walker in position $x$ at time $t$ given the model and given all previous positions $X_{t-1}$, i.e. $P_{\mathrm M}\left(x_t \mid X_{t-1}, \theta \right)$.
The likelihood of a sequence of $t$ events \citep{Schuett2017} is therefore given by the product of $t$ conditional probabilities ${\cal L}_M(\theta|X_t)$, which is given by  
\begin{equation}
    \label{p3:eq:l_m_xn}
    \mathcal{L}\left(\theta \mid X_n\right) = P_{\mathrm M}\left(x_1 \mid \theta\right) \prod_{i=2}^n P_{\mathrm M}\left(x_t \mid X_{t-1}, \theta \right) \;.
\end{equation} 

In order to estimate the parameters of the model, we use Bayes' theorem to compute the probability of the parameters $\theta = [\gamma, r_i, r_j, \phi, \lambda]$ given the data as
\begin{equation}
\label{eq:Bayes}
    P(\theta \mid X_n) = \frac{L_{\mathrm M}(\theta \mid {X_n})P(\theta)}{\int_\Theta L_{\mathrm M}(\theta \mid X_n)P(\theta){\mathrm d}\theta} \;.
\end{equation}

A large literature exists to solve likelihood-based parameter estimation computationally. 
In order to leverage the full power of the likelihood-based approach we estimate the full Bayesian posteriors for each parameter using a differential evolution adaptive metropolis sampling algorithm (DREAM) \citep{Laloy2012, Shockley2018}. More details are provided in the Methods section.

\subsection*{Parameter estimation results}
We estimated the values of the free parameters of the SAW model for each subject independently based on data. The chosen priors (see Appendix for details) were relatively uninformative truncated Gaussians. Figure \ref{fig:result} presents the marginal posteriors for each participant in the study. In the Appendix (Table \ref{tab:params}) we present the point estimates and 98\% confidence intervals \citep{Kruschke2014} for each parameter.

\begin{figure}
    \centering
    \unitlength1mm
    \begin{picture}(\linewidth,100)
        \put(-18,0){\includegraphics[width=1.2\textwidth]{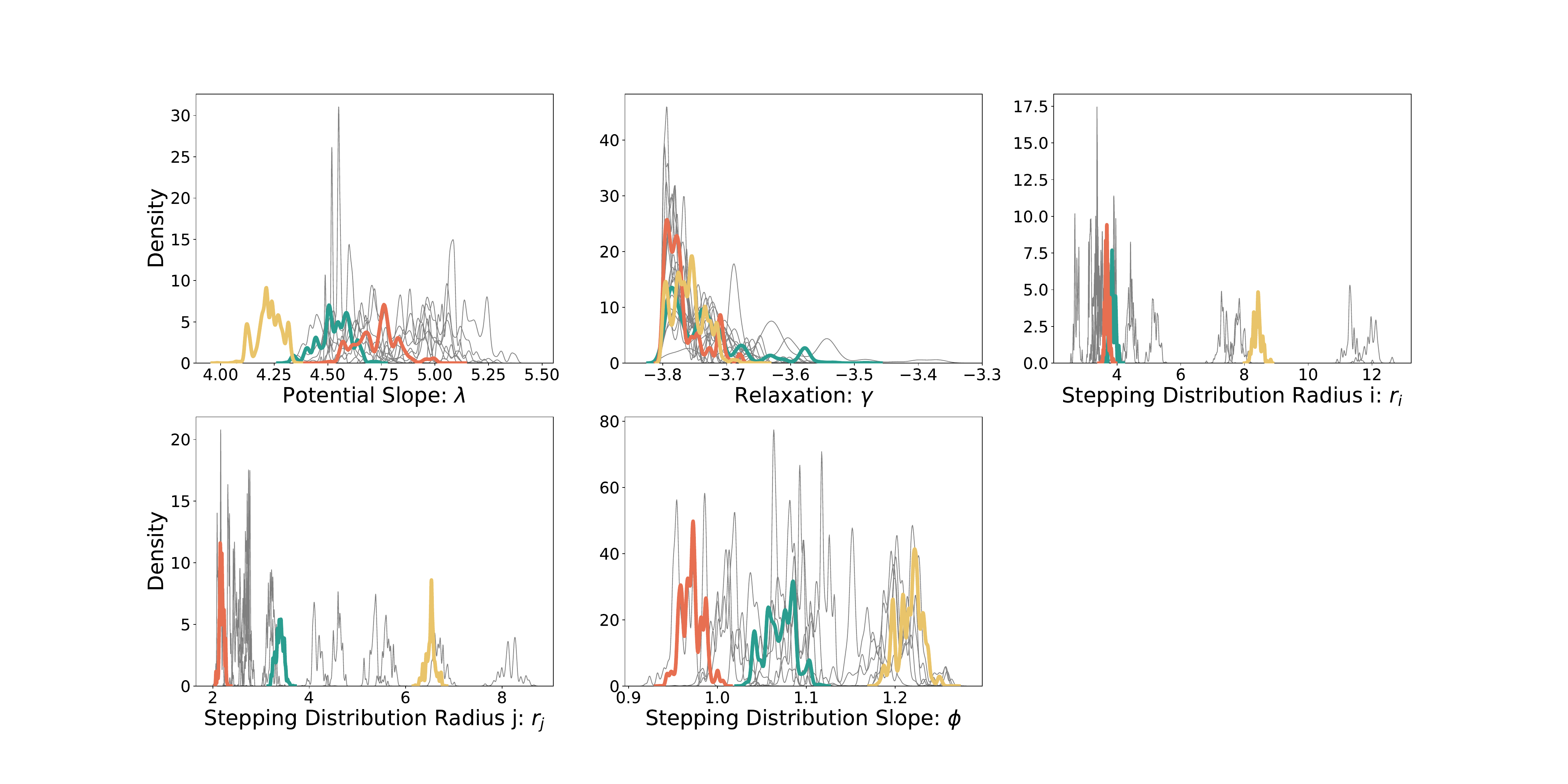}}
        \put(0,6){(A)}
        
        \put(123,7){\includegraphics[width=0.35\textwidth]{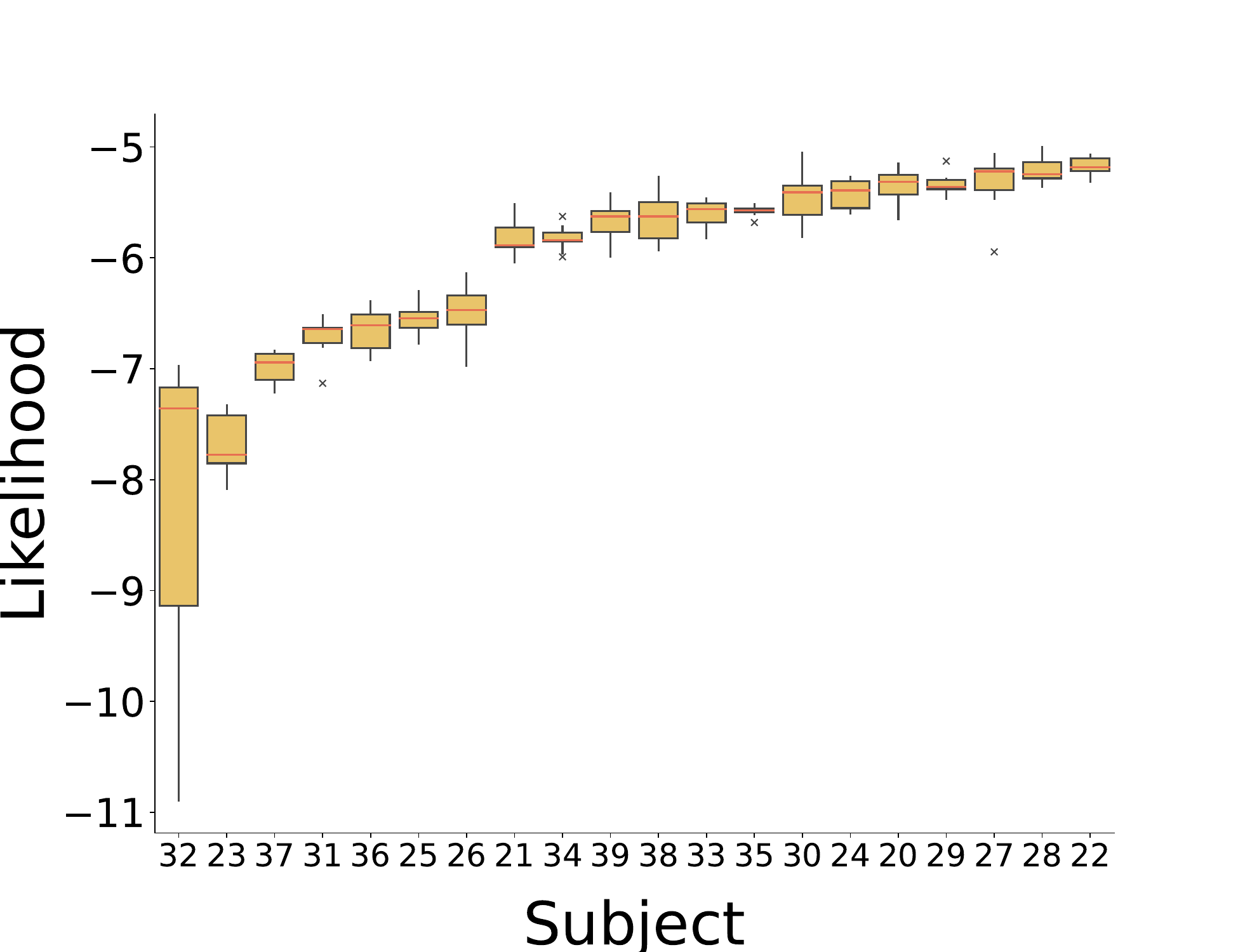}}
        \put(123,5){(B)}

    \end{picture}
    \caption{\label{fig:result}
        (A) Marginal posteriors for the five estimated parameters; the grey lines represent different participants, colored lines highlight three participants for comparison. For all five parameters, the posteriors converged to distinct peaks for the different participants.  (B) A box plot based on the sum log likelihood of all trials, sorted by mean log likelihood across participants.
    }
\end{figure}

\begin{figure}
    \centering
    \unitlength1mm
    \begin{picture}(180,60)
        \put(0,0){\includegraphics[width=0.35\textwidth]{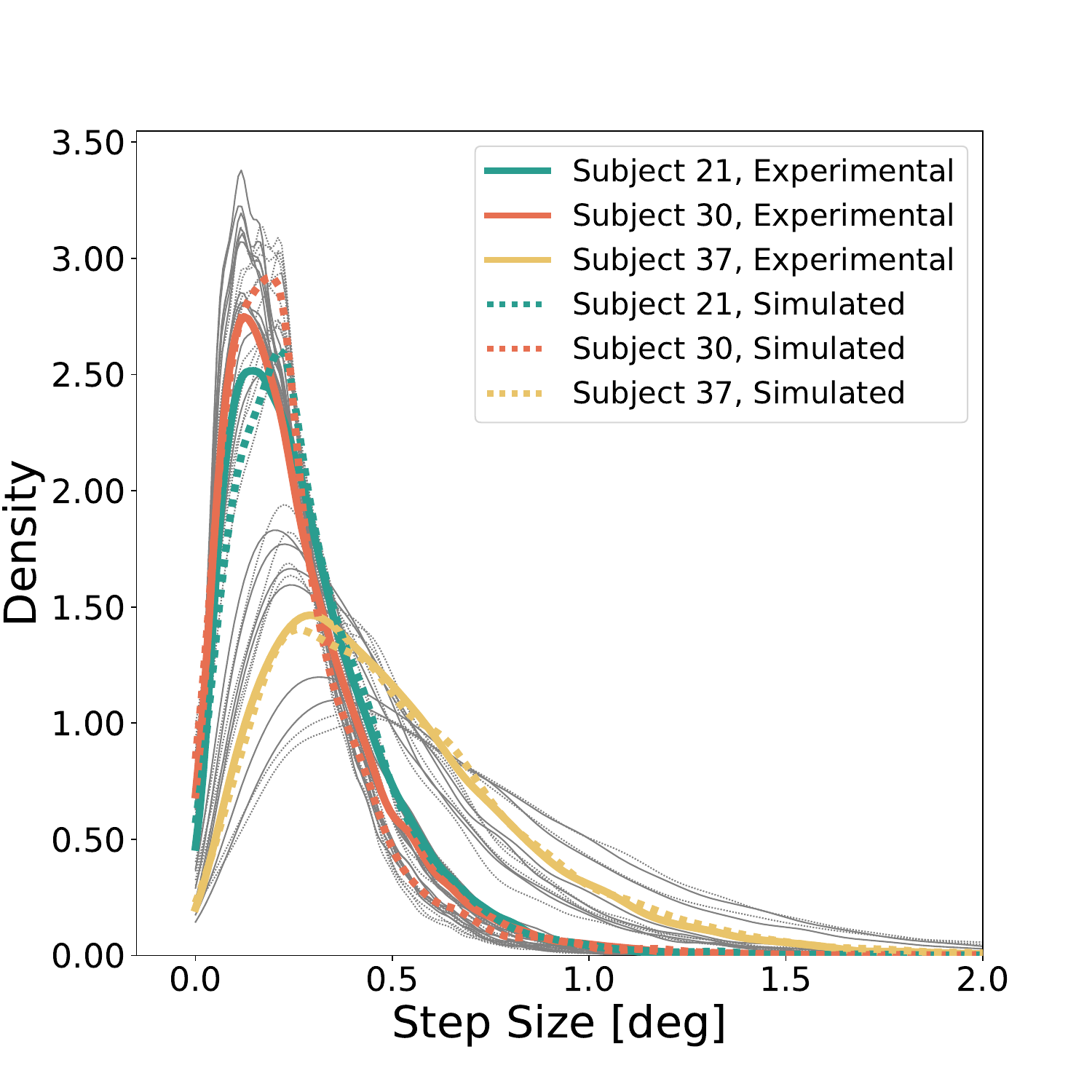}}
        \put(0,0){(A)}
        \put(60,0){\includegraphics[width=0.35\textwidth]{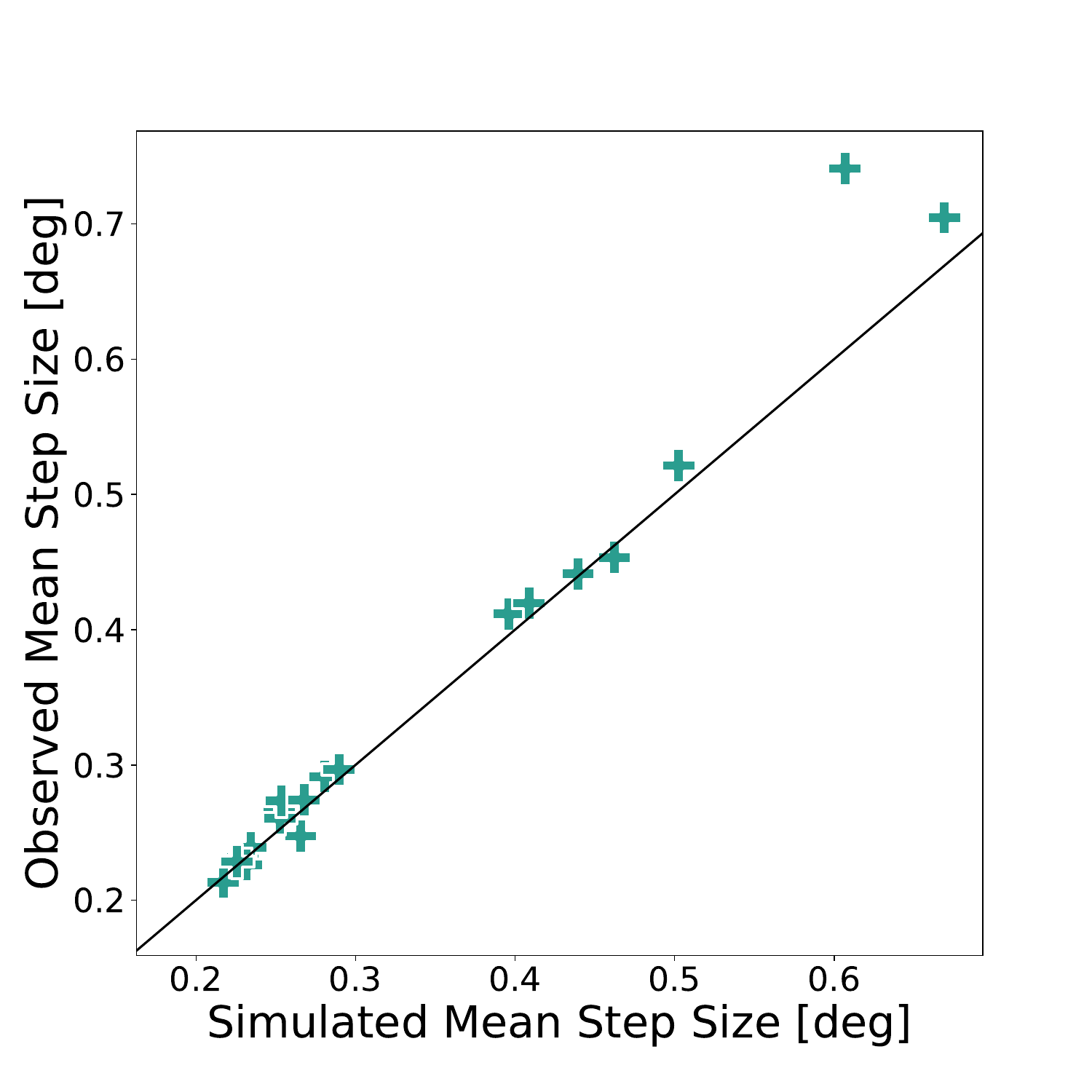}}
        \put(60,0){(B)}
        \put(120,0){\includegraphics[width=0.35\textwidth]{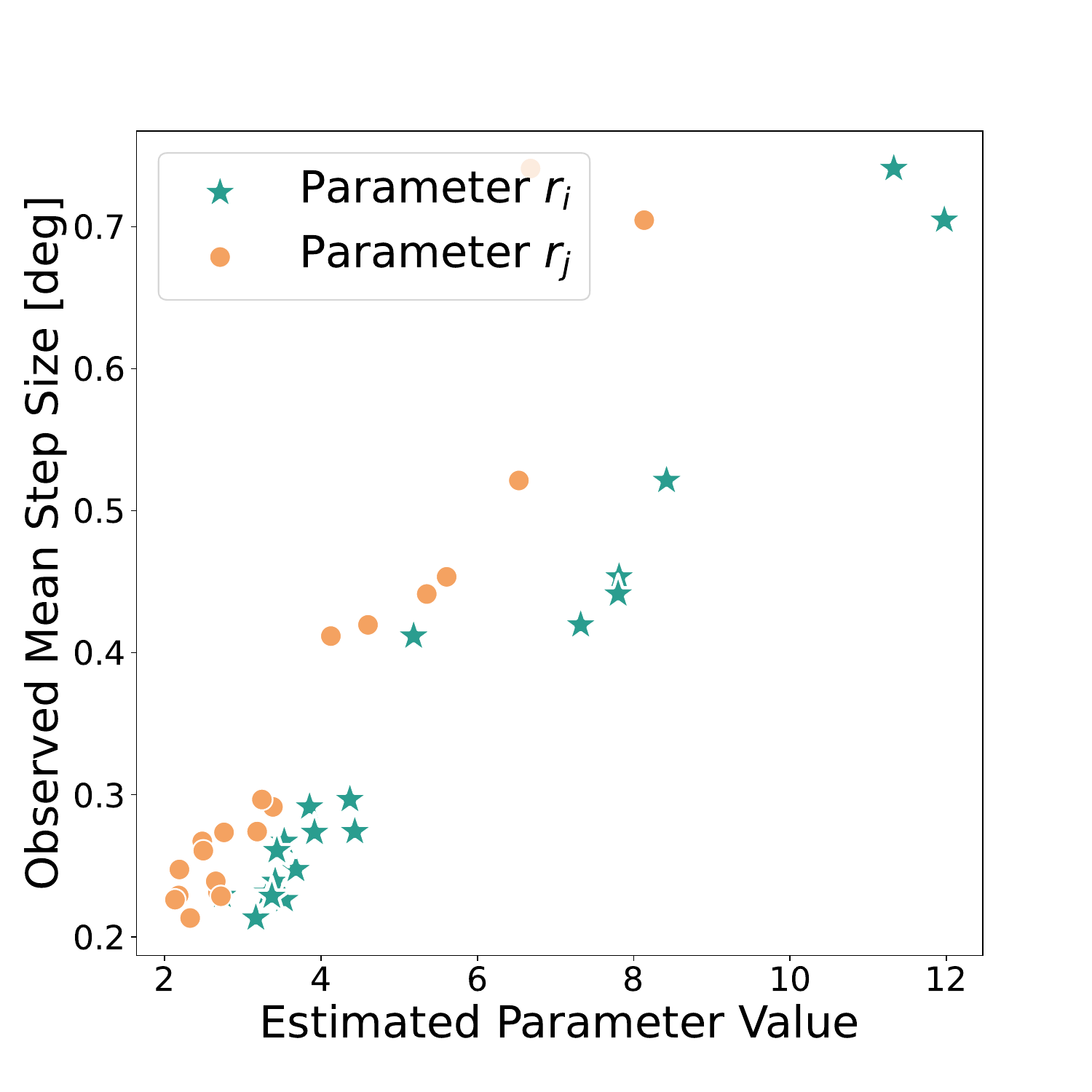}}
        \put(120,0){(C)}

    \end{picture}
    \caption{\label{fig:step_rij} The fit between step sizes in the model and in the data. (A) shows the step size distribution, where highlighted lines represent individual subjects. (B) shows the correlation between simulated and empirical step sizes. (C) shows the  correlation between the observed step size and the parameter value of $r_i$ and $r_j$. As expected, the two parameters correspond directly to the stepping distribution  in the data.}
\end{figure}
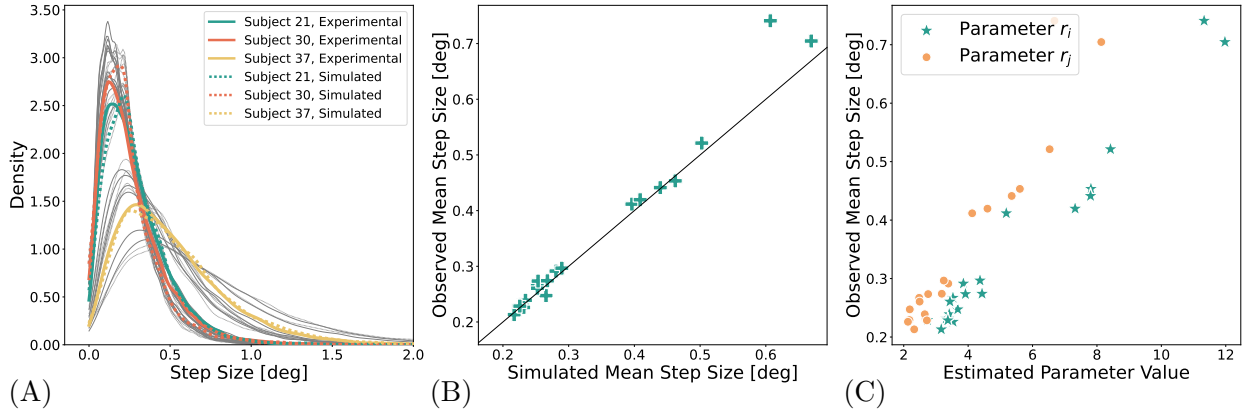

Biologically plausible models are designed to be grounded in real-world biological mechanisms and processes. As such, the values of their parameters reflect the known properties of these mechanisms. It is therefore informative to investigate the parameter values themselves, as they permit inferences about concrete aspects of the data generating process. First, the parameters $r_i$ and $r_j$ correspond to the widths of the stepping distribution.
The final selection map in the model (Eq. \ref{eq:stepping}), and consequently the resulting predictions for step sizes, depend on the stepping distribution containing $r_i$ and $r_j$, as well as the confining potential and activation. We define the step size as the distance travelled from one measured experimental sample to the next, i.e., 2~ms as we are using a sampling rate of 500~Hz. As shown in Figure \ref{fig:step_rij}C the parameters $r_i$ and $r_j$ are strongly correlated with the empirical step size distribution. 

The parameter $\phi$ represents the slope of the stepping distribution. Higher values are associated with a stronger tendency to move along the cardinal directions. The estimated values for $\phi$ range from 0.9 to 1.3, indicating that the preference for cardinal directions is stronger in some participants than in others. The model captures individual differences in the stepping distributions, as demonstrated by the distinct posteriors obtained for $\phi$, $r_i$ and $r_j$.

Next, the parameter $\gamma$ relates to the speed of memory decay in the process, where smaller, more strongly negative values cause slower decay, i.e., longer memory of the visited locations, and larger values closer to 0 cause faster memory decay, i.e., shorter memory. It was bounded in the estimation to a minimum of $-4$, as smaller value, i.e. less relaxation, caused the activation in the system to continuously increase,  making the model numerically unstable. We find an average value of $\gamma$ of 3.75 which indicates that activation decreases by 25\% over the course of a trial duration of 3~s. Parameter $\gamma$ accounts for relatively small individual variability. For most of the participants, estimates indicate a long memory for activation that represented the past trajectory in the system.

Finally, parameter $\lambda$ represents the slope of the confining potential. The shape parameter of the potential function was fixed to 3, fixing the qualitative form to a steeper gradient than in a quadratic case \citep{Engbert2011}. The slope was considered a free parameter for the estimation. As the different participants' data lead to different decay and stepping parameters, the slope of the potential needs to accommodate different resulting values of mean system activation.

\subsection*{Posterior predictive checks}
Using the estimated parameters (i.e., the posterior parameter distributions), we simulated artificial data sets of the same size as the experimental test data sets. We know that the model can, in principle, recreate statistical tendencies of ocular drift found in experimental data on a qualitative level \citep{Engbert2011}. Posterior predictive checks show that the fitted model reproduces the expected statistical tendencies, i.e., turning angle- and step size distributions, including inter-individual differences.

Fitted primarily by parameters $r_i$ and $r_j$, the step size distribution is represented in Figure \ref{fig:step_rij} A. As defined above, the step size is the spatial distance between two subsequent samples. In order to condense the distribution of step sizes into one summary value, we computed the mean of each distribution. Thus, Figure \ref{fig:step_rij} B shows the correspondence between the mean true and simulated step sizes. The model fits the step size very precisely and perfectly captures the differences in the individual preferred step size. 

\begin{figure}[p]
    \centering
    \unitlength1mm
    \begin{picture}(150,120)

        \put(0,80){\includegraphics[width=0.45\textwidth]{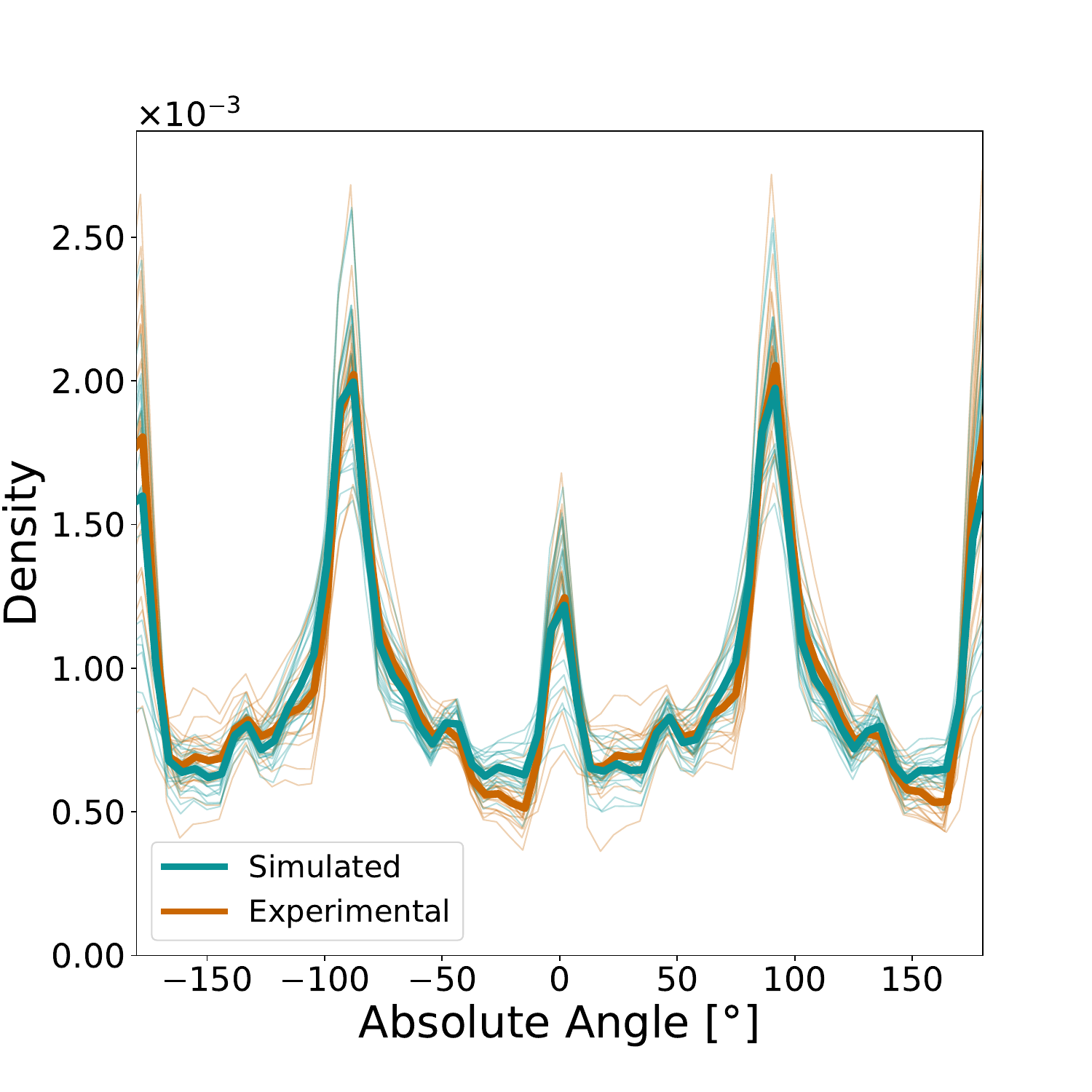}}
        \put(0,80){(A)}
        \put(80,80){\includegraphics[width=0.45\textwidth]{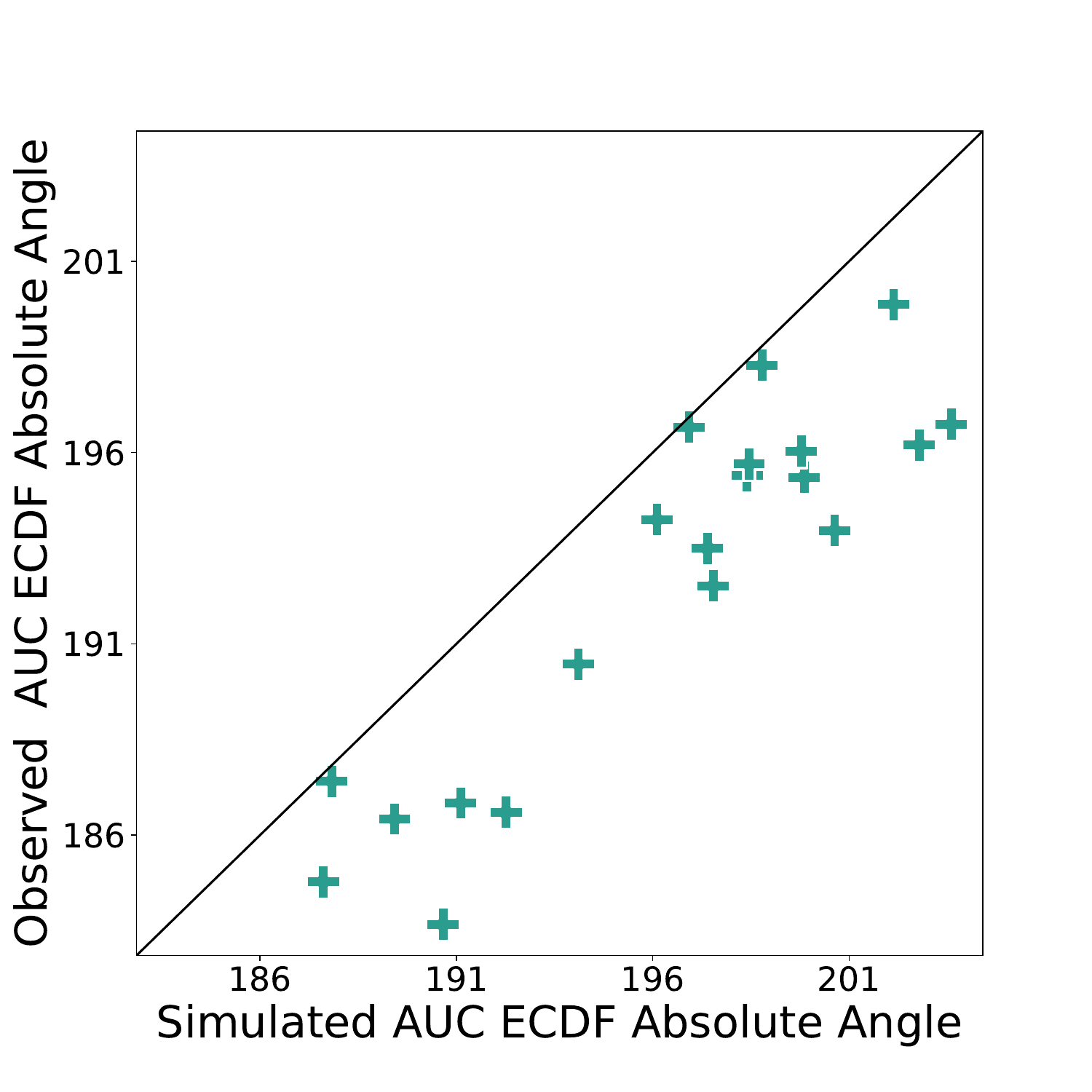}}
        \put(80,80){(B)}

        \put(0,0){\includegraphics[width=0.45\textwidth]{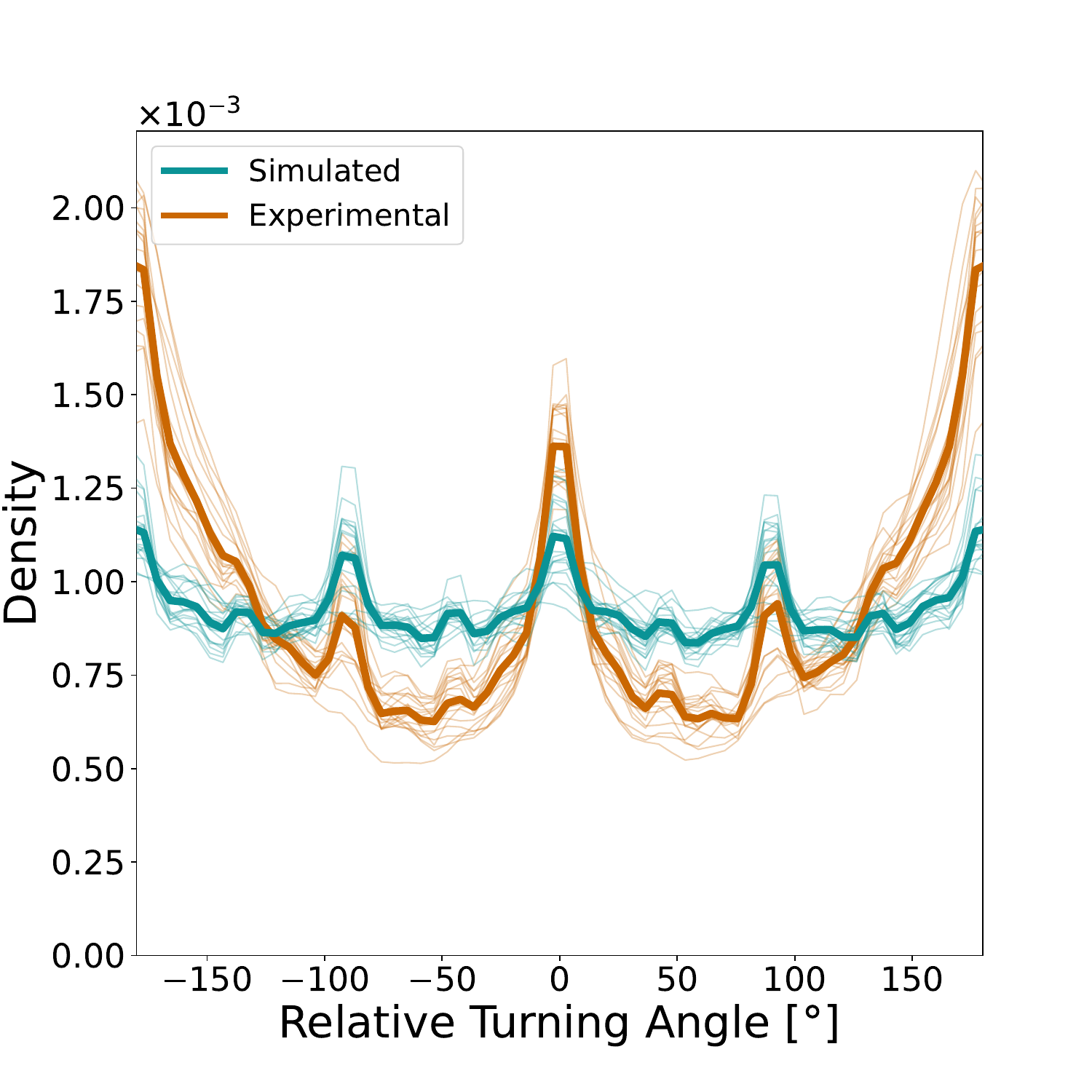}}
        \put(0,0){(C)}

        \put(80,0){\includegraphics[width=0.45\textwidth]{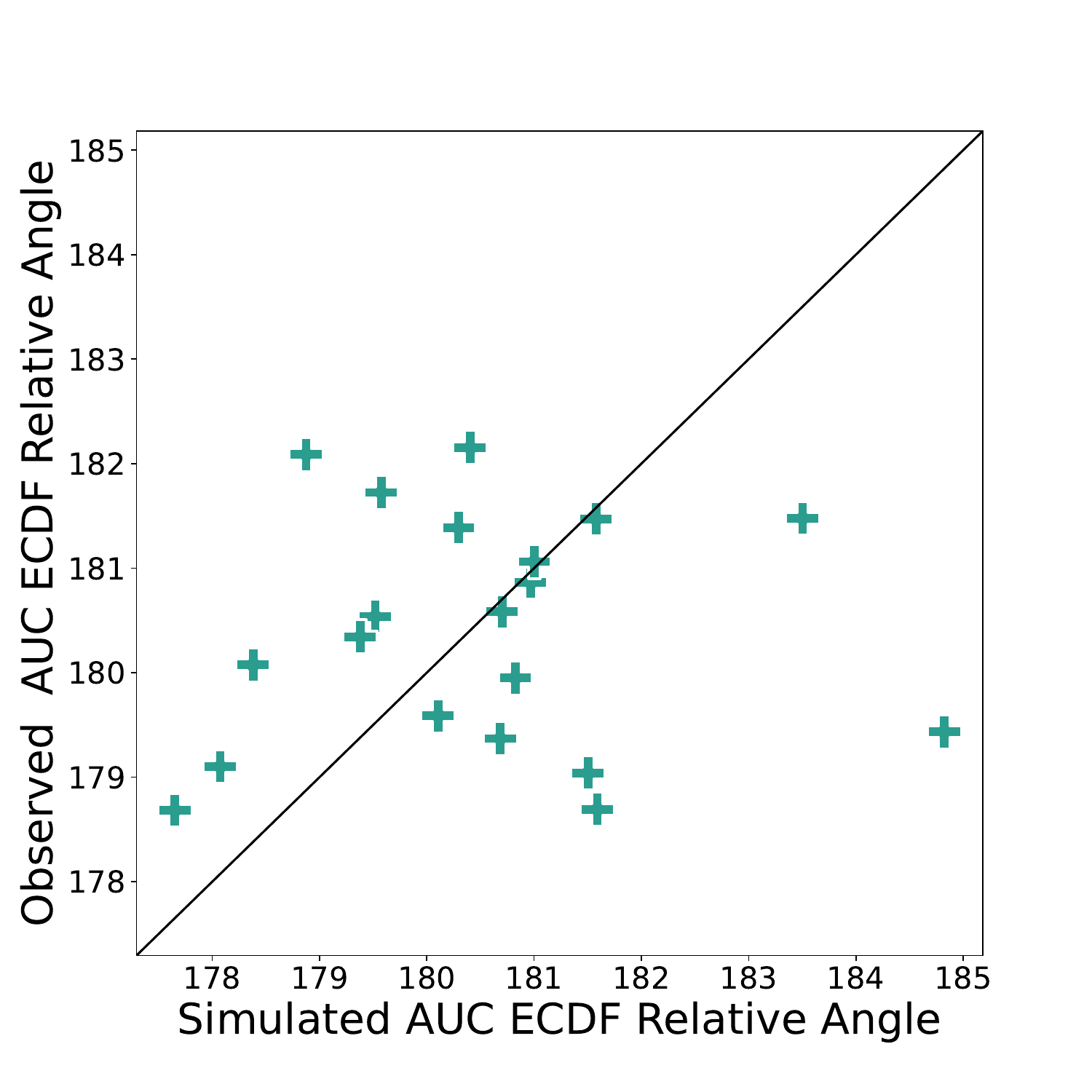}}
        \put(80,0){(D)}
    \end{picture}
    
    \caption{\label{fig:angle} Turning angle distributions. (A) and (C) show the absolute and relative turning angles of one sample to the next. Thin lines represent individual subjects and thicker lines represent the means. Simulated data is shown in green while experimental data is red. Panels (B) and (D) show the respective correlations between simulated and experimental data. We characterize the angle distributions using the area under curve (AUC) of the empirical cumulative density function (ECDF). For details see the Methods section.}
\end{figure}
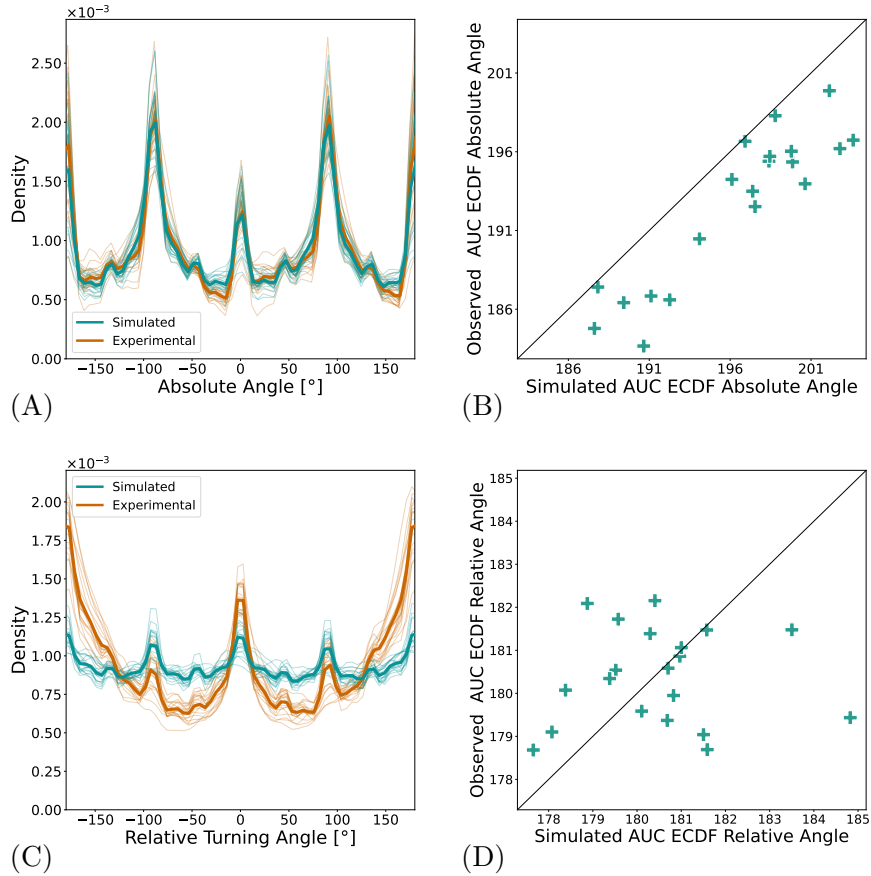

Next, we investigated the absolute and relative turning angles. There exists a preference in both ocular drift and (micro)saccades for movements in the cardinal directions, which may be caused by the structure of the ocular muscles \citep{Sparks2002}. This fact is captured well by the model (Figure \ref{fig:angle}A). The relevant parameter responsible for this effect is the stepping distribution slope $\phi$, which shapes the stepping distribution to have a stronger or weaker preference for the cardinal directions. In order to ascertain whether the differences in individual behavior can also be reproduced, we use the area under the cumulative density function as a summary statistic (see the Methods section). The result in Figure \ref{fig:angle}B shows that the model is capturing individual differences. Furthermore, there is also a tendency for movements to be in line with or orthogonal to the previous movement vector. While this tendency is a lot more pronounced in the experimental data than in the simulated data, the simulations do show a qualitative reproduction of the trend (Figure \ref{fig:angle}C). However, it is likely that the peaks are driven by the preference in absolute angles, as no mechanism for relative angles was built into the model. We propose that the difference between the simulated and experimental relative turning angle distributions reveals the part of the relative turning angle distribution that must be accounted for by an additional, independent model mechanism.

\begin{figure}
    \centering
    \unitlength1mm
    \begin{picture}(\linewidth,60)
        \put(0,0){\includegraphics[width=0.35\textwidth]{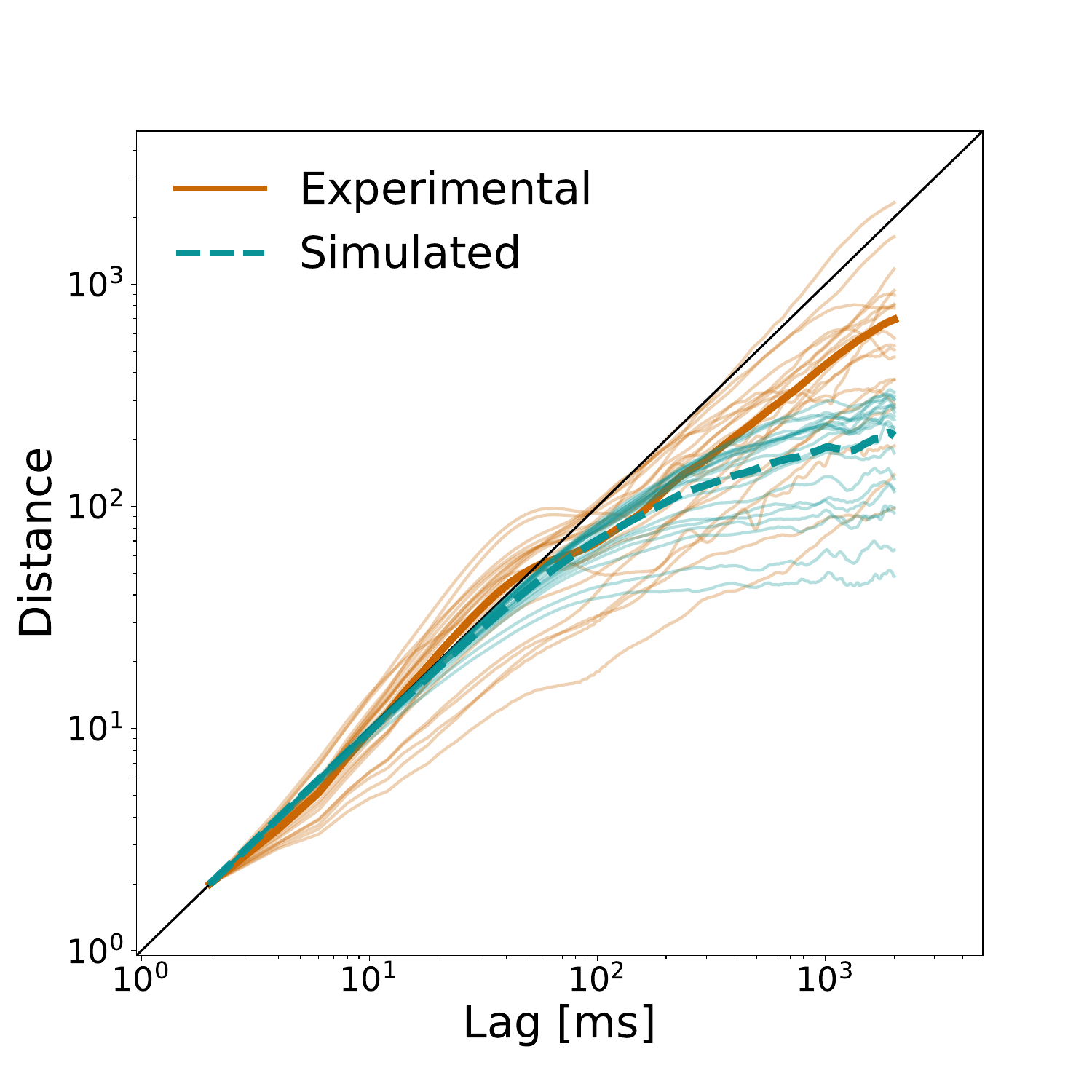}}
        \put(0,0){(A)}

        \put(60,0){\includegraphics[width=0.35\textwidth]{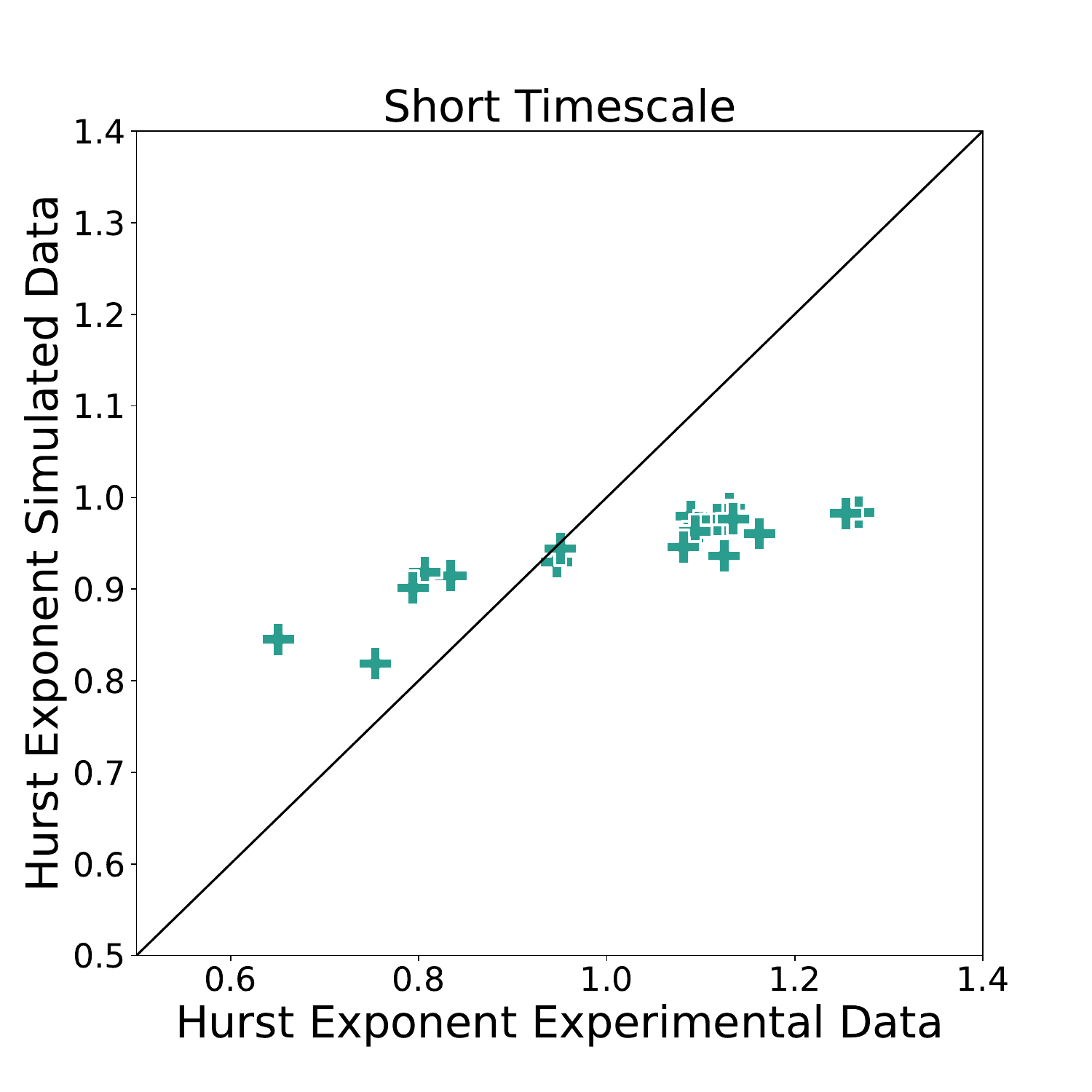}}
        \put(60,0){(B)}

        \put(120,0){\includegraphics[width=0.35\textwidth]{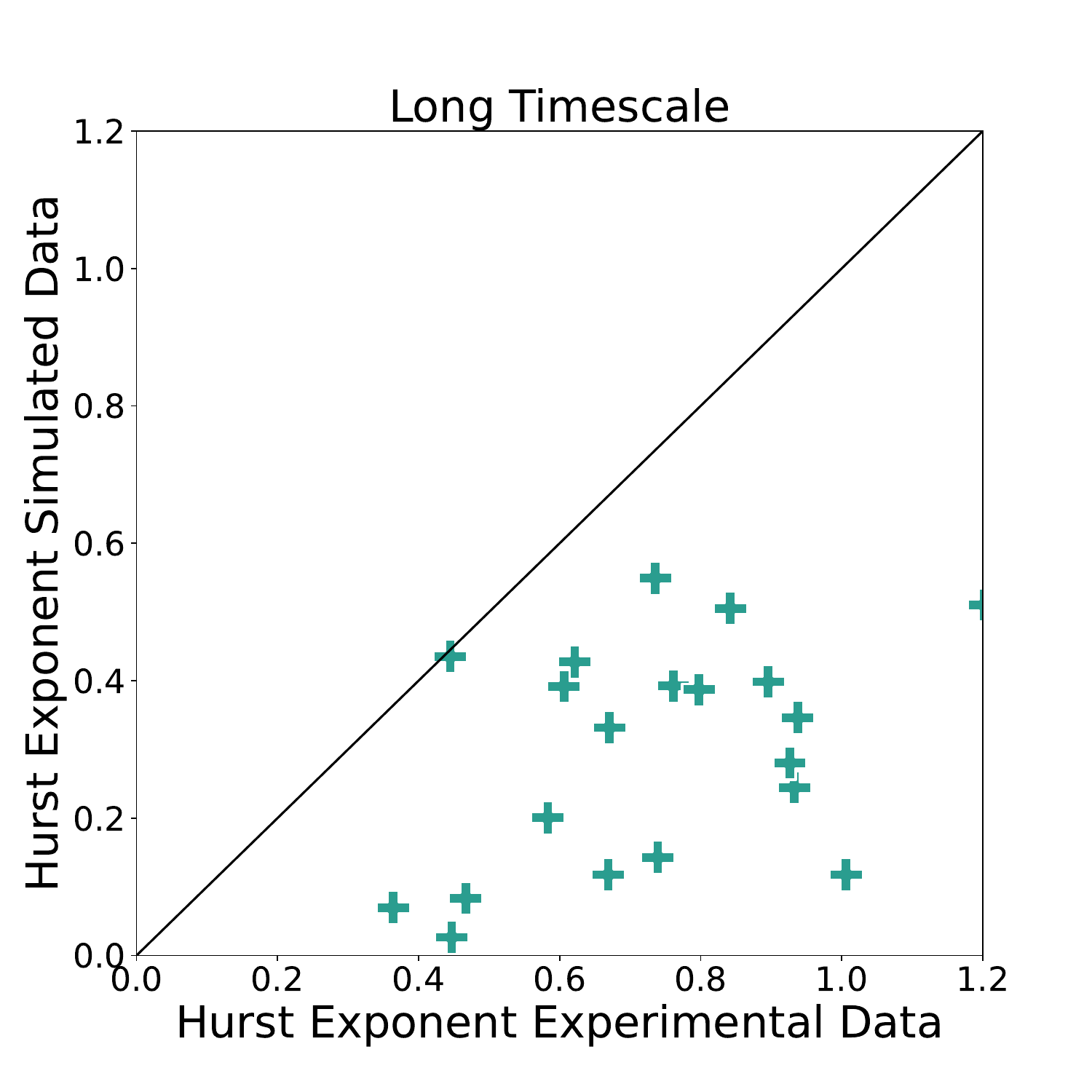}}
        \put(120,0){(C)}

    \end{picture}
    
    \caption{\label{fig:msqd} Persistent and Anti-persistent behavior. (A) shows the MSD of simulated and experimental data, where thinner lines represent individual subjects and thicker line represent the averages of experimental data in red and simulated data in green. Note that the lag is given in ms, whereas the distance is given in normalized units, in order to visualize the slopes in comparison to the identity line. (B) and (C) shows the correlation of simulated and experimental data of linear fits of the Hurst Exponents for the short and long timescale for individual subjects.}
\end{figure}

Empirical ocular drift data, has been found to be persistent at small time lags and anti-persistent at longer timescales. As shown in Figure \ref{fig:msqd}A, the persistent trend in our data is not as pronounced for all participants as in comparable previous studies \citep{Engbert2004,Herrmann2017}. The anti-persistent period begins around 60 to 80~ms after stimulus onset. The data simulated using the fitted parameters reflects this behavioral change well. However, it behaves more randomly than truly anti-persistent at short timescales and becomes too strongly persistent at long timescales. Experience with the model shows that it can produce truly anti-persistent behavior by varying the free parameters. Possible reasons why this trend was insufficiently captured by the fitted models are that the MSD is a less dominant tendency compared to other statistics and that our selection of free model parameters were correlated in a way that limited the ability to fit this particular tendency. Alternatively, the self-avoidance of the model may not be the only cause for early persistence, suggesting a model with an explicit exploration mechanism. Nonetheless the qualitative  change in behavior is clearly present in the simulated data. Accordingly, the correlation between the experimental data and the simulated data show that the present model fits only account for a small part (roughly 10\%) of the individual variation (\ref{fig:msqd}B and C).

\subsection*{Investigating microsaccades}
Previous work concerning the SAW model has suggested a connection between the self-avoiding properties of the random walk and microsaccade triggers. A reduction in movement tends to precede microsaccades \citep{Engbert2006}.  Following this reasoning, in the model, a decrease in movement corresponds to a build-up of activation in the current position.  Thus, we investigated whether the activation predicted by the model is indeed related to the occurrence of microsaccades. Specifically, we calculated the activation $q_t(i,j)$ at times relative to microsaccade onsets $t_{\textrm MS-on}$ and offsets $t_{\textrm MS-off}$ using the experimental data.

\begin{figure}
    \centering
    \includegraphics[width=1.0\linewidth]{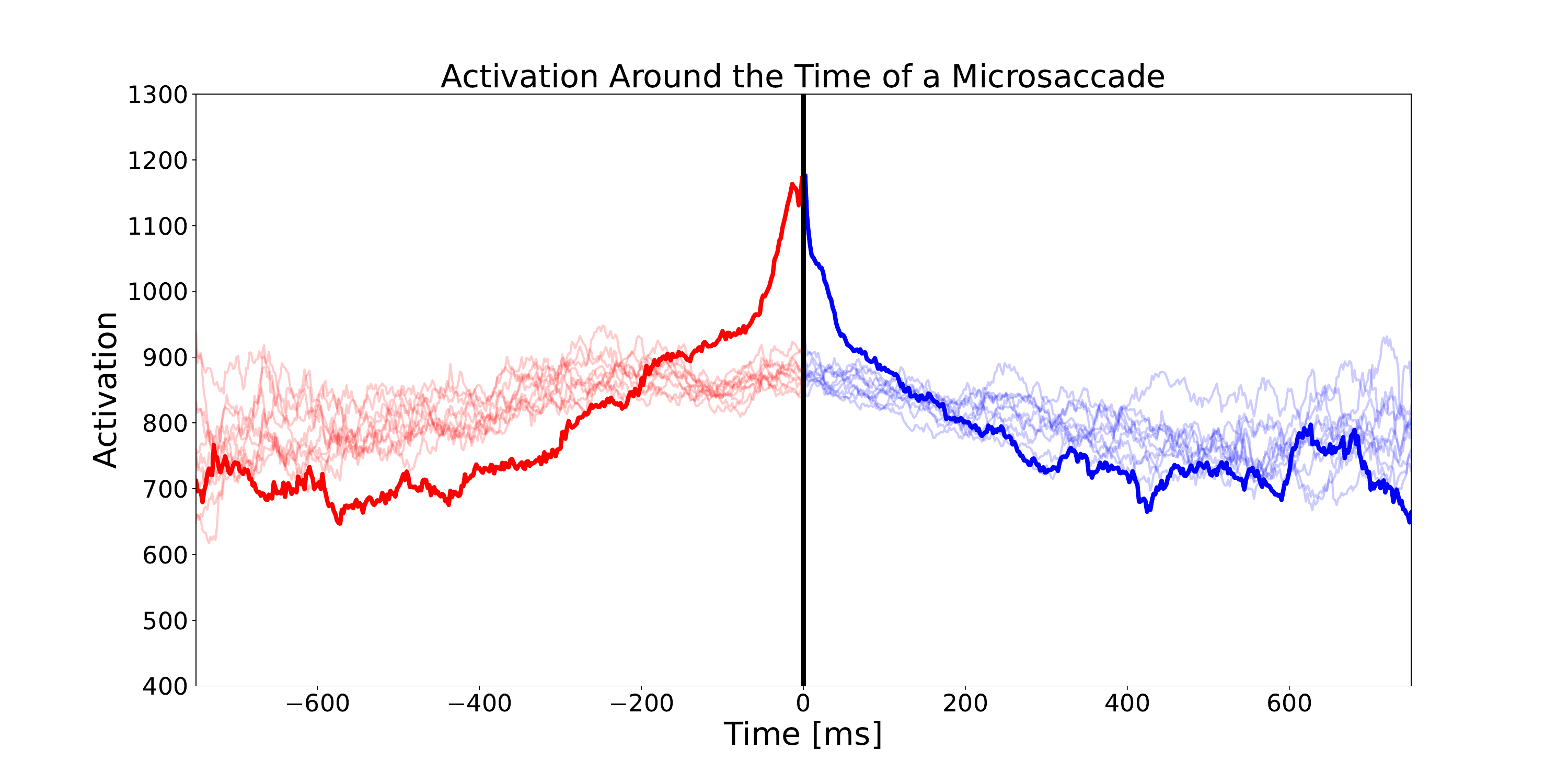}
    \caption{\label{fig:sactime} The average activation in the model around the time of a microsaccade. The less opaque lines represent randomized controls, where microsaccade onsets were randomized over the trials within one subject. In the data we find a distinct rise in activation before the time of a microsaccade and a drop in activation after.
    }
\end{figure}

Figure \ref{fig:sactime} shows the average activation around the time of a microsaccade, which is consistent with the hypothesis that high activity levels are related to triggering microsaccades. To better understand the extent of the effect, we randomized the microsaccade onsets within each subject and computed the same trajectory on the randomized data. The activation rises more before a microsaccade and drops more steeply than the randomized controls.

\subsection*{Model comparisons}
The proposed model comprises three main components: the random walk with a stepping distribution, the self-activated trace memory, and the potential. In theory, the combination of both creates an interplay between persistence and fixation control. In order to better understand the role of each component, we created 3 control models by removing individual components. Specifically we investigate 
\begin{enumerate}
    \item the full baseline SAW model, that contains all three components
    \item a model that is a random walk in a potential (W),
    \item a model that is a random walk without a potential(W-NP), 
    \item a model that is a random walk with self-avoidance but without a potential (SAW-NP).
\end{enumerate}

Note that the latter two models differ from the first two significantly in that they are not generative and, therefore, are not biologically plausible. In the absence of a potential it is still possible to compute the likelihood, but it is not possible to usefully simulate data from them, as there is nothing stopping the walker from simply walking away. We include them here, because they provide a relevant comparison, however, these models must be treated as substantially different concerning the conclusions they permit.

\begin{figure}
    \centering
    \unitlength1mm
    \begin{picture}(\linewidth,140)
        \put(30,80){\includegraphics[width=0.8\textwidth]{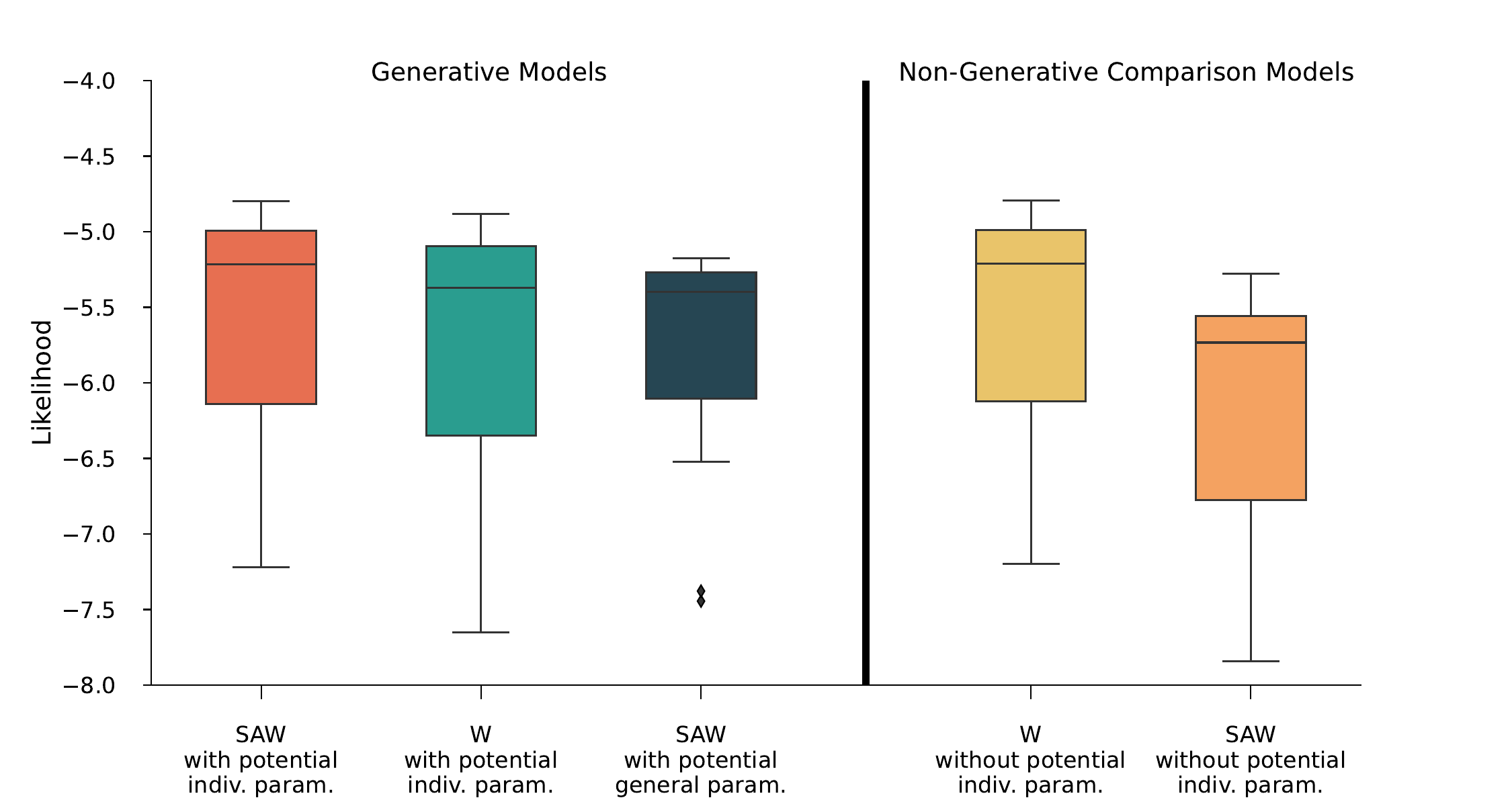}}
        \put(25,80){(A)}
                
        \put(27,0){\includegraphics[width=0.8\textwidth]{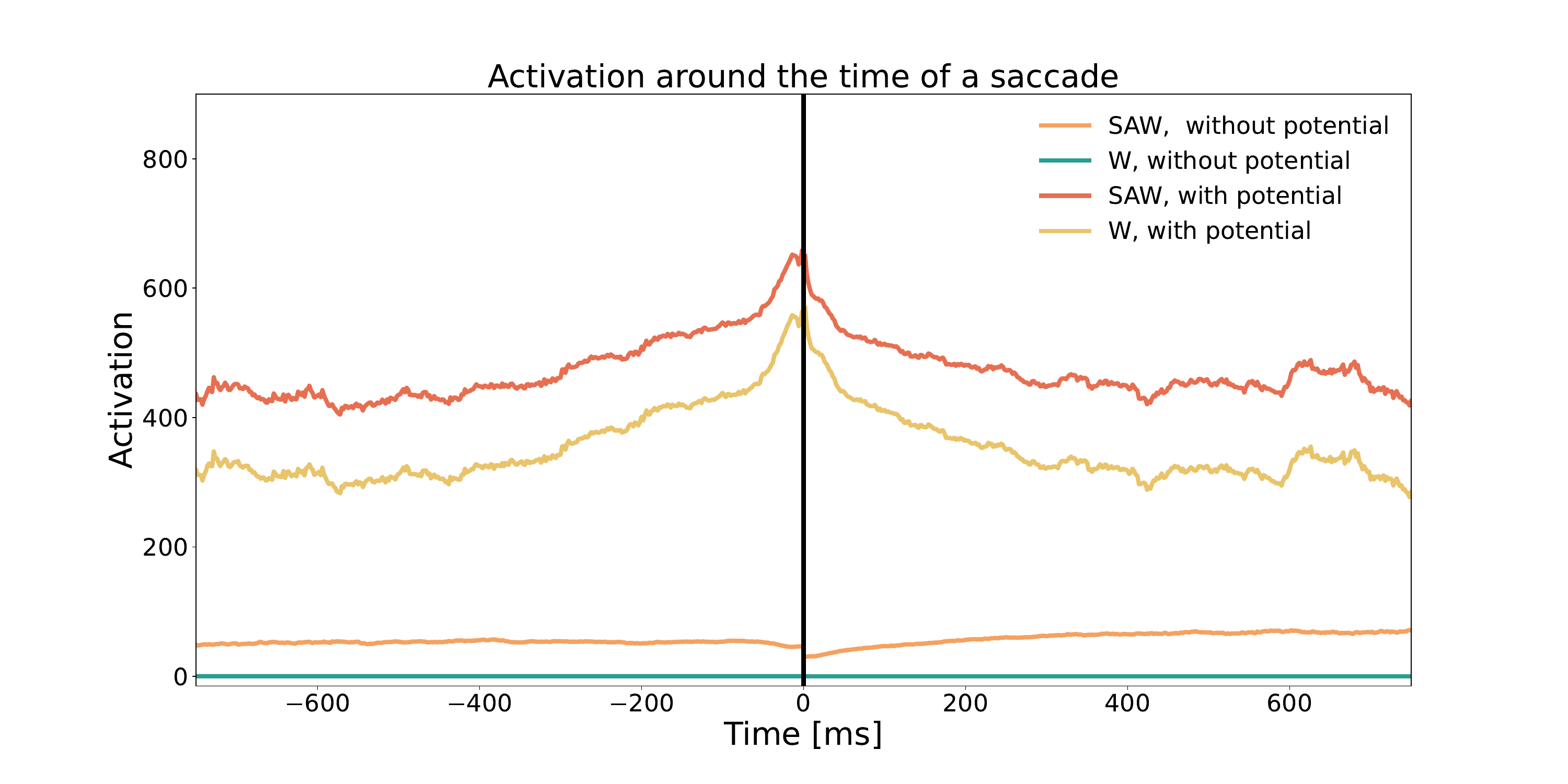}}
        \put(25,5){(B)}
    \end{picture}
    \caption{\label{fig:modelcomp} Comparisons of the different model versions. Panel A shows the 5 different versions of the model, including different components. We show that among biologically plausible models that the proposed mechanisms and the individual fitting procedure confer a likelihood benefit. Among Non generative models a pure random walk with fitted step sizes, while not biologically plausible, also achieves a high likelihood score. Panel C shows the model activation around the time of a saccade. The activation peak emerges only in the models that contain a confining potential.}
\end{figure}

First, we compare the models in terms of their likelihood (Figure \ref{fig:modelcomp}A). Each model was evaluated on the test data set, using the same parameters wherever applicable. Our findings indicate that SAW outperforms the model version without self-activation (W). Removing the potential and maintaining self-avoidance (SAW-NP) also reduces performance. However, we also find that the non-generative model without either potential or self-avoidance (W-NP) produces an almost identical likelihood to SAW. This result suggests that individually fitted stepping distributions and general random walk behavior capture the data's most predominant features. It is important to note that the likelihood is a very general measure of model performance -- the model (W-NP) is neither biologically plausible nor can it capture any additional statistical properties of the data. Thus, the lesson learned from this modeling study is twofold: First, a high likelihood is not always a guarantee of an appropriate model \citep[see also][]{Schuett2017}, and second, that either a potential or self-avoidance is not beneficial to model performance. Each component, added individually, actually reduces the model likelihood. The fact that their joint effect reestablishes a similar likelihood while simultaneously making the model more biologically plausible should be considered a success. We added another model to this comparison: the SAW model without individual parameter fits by subject. We observe that on average, fitting individual subjects is a distinct benefit. However, the averaged parameter model reduces high likelihood values and the number of very low likelihood values. 

Second, we used our comparison models to investigate which components drive the microsaccade effect. The apparent picture in Figure \ref{fig:modelcomp} shows that the activation peak is present only for the two models with a potential. This finding indicates that the effect is not driven, as we supposed, by a build-up of self-generated activation but rather by the potential. This result is consistent with the idea that the microsaccades that drive the effect are related to the control of the fixation position.

\section*{Discussion}
Fixational eye movements display significant randomness, and their origin and purpose have been debated in the literature. Mathematical modeling approaches have contributed insights into the neurophysiological origin of the movement \citep{Eizenman1985, BenShushan2022}, the desirable or undesirable consequences of the motion on image processing \citep{Schmittwilken2022, Anderson2020}, and the spatiotemporal statistics of the drift trajectories \citep{Burak2010, Engbert2011, Roberts2013}. We performed Bayesian likelihood-based parameter inference of a self-avoiding random walk model for fixational drift at the level of individual observers. The estimation of the parameters converges to distinct marginal posteriors, and data simulated based on the fitted models reproduces individually characteristic behavior. In the second step, we propose a relationship between the microsaccade rate and peaks in the model's latent activation state. An exploratory, data-driven analysis confirms this intuition.

\subsection*{Individual variability}
A complex combination of factors, such as oculomotor control, attention, and cognition, controls fixational eye movements. The specific observed patterns vary significantly by individual both for measures of ocular drift \citep{Cherici2012} and microsaccades \citep{Poynter2013}. Our results indicate that the individual variability in drift can at least partly be captured by the parameters of the SAW model. The average preferred step size is a particularly pertinent example ($r_i, r_j$), but also directional preferences ($\phi$) and potential slope ($\lambda$) are different between subjects. These parametrizations are sufficient to simulate data that mirrors the characteristic features of individuals. To our knowledge, this is the first paper to model individual variation of fixational eye movement trajectories.

The variability in drift between individuals is related to the individual variability in acuity \citep{Clarke2021}. Individual differences in fixational eye movement may be related to a range of factors, including the precise acuity of the eye and the tendency to maintain precise fixation \citep{Cherici2012}. Moreover, attentional preferences found in macroscopic eye movement during facial feature viewing translate to microscopic eye movement preferences \citep{Shelchkova2019}.

We find a distinct benefit from using individually fitted data sets over a single averaged value for each parameter. This benefit becomes apparent in better convergence properties of the parameter estimation, as the solid and distinct influences otherwise cause complex, multi-modal posteriors. Moreover, the individual fits cause a higher overall likelihood and fit of individual characteristics. However, while model based in averaged parameter values has a lower likelihood, it also reduces the number of very low likelihood data points. Thus, individuals whose data can not be easily fitted would benefit from a normative influence of other subjects. This observation suggests using a hierarchical Bayesian modeling approach in future studies.

\subsection*{The confining potential}
The characteristic Mean Square Displacement (MSD) of fixational drift is persistent at small time lags and anti-persistent at longer lags \citep{Engbert2004,Herrmann2017}. This is consistently the case, when averaging over large amounts of data. On an individual level, however, this tendency is not always equally pronounced. While the SAW model reproduces the transition to anti-persistent behavior well, it does not adequately represent the persistent component, with the current data. As, in principle, self-avoiding random walk models can produce persistent behavior \citep{Engbert2011, Roberts2013},  this may be caused by a number of factors including the selection of fitted free model parameters, the relatively low persistence in the present data set, or a dominance of other statistical tendencies. Alternatively, it is possible that the strength of the persistent trend, which the SAW model frames as the result of self-avoidance, is in fact amplified by an additional explicit exploration mechanism which remains to be identified. 

Exploration, or the explicit persistence of the trajectory, is highly variable, even between trials. The confining potential in the SAW model is static, representing a fixed intended fixation position. These assumptions are a simplification, as the intention may change over time. Experimentally, we find a large amount of variation in the cohesion of the drift. In some trials, drift consistently occurs around a specific position. In others, it is evident that two intended fixation positions coincided over the course of the trial. In others, drift consistently maintains its direction away from the starting point. As the task and stimulus in the experiment were the same for all trials, there is little evidence to explain this variation, aside from random variation or influence of recent past stimuli. A potential future direction for the model could be implementing a dynamic confining potential centered around a moving average of several recent samples. The limitation underscores the need for more comprehensive and accurate models to capture better the complex and individual characteristics of ocular drift behavior.

\subsection*{The relationship between drift and microsaccades}
Early hypotheses suggested that microsaccades serve a corrective function for ocular drift \citep{Ditchburn1952, Cornsweet1956, Nachmias1959}. Experimentally, however, no reliable correlation data confirms this hypothesis, as microsaccades can be explorative as well as corrective. Specifically, at shorter time scales, microsaccades induce persistent correlations, while at longer time sc
ales, they tend to reverse the movement to correct the fixation position \citep{Engbert2004}. Additionally, studies have shown that during high-acuity observational tasks, participants naturally suppress microsaccades without training \citep{Bridgeman1980,Winterson1976}, leading to the conclusion that microsaccades may serve no useful purpose \citep{Kowler1980}. However, more recent research has demonstrated that microsaccades can enhance the visibility of peripheral stimuli \citep{MartinezConde2006}, facilitate high acuity vision \citep{Poletti2013, Intoy2020}, and are responsive to task demands \citep{Ko2010}, suggesting a direct link between microsaccade activity and visual perception. 

Thus, the role of microsaccades in fixational eye movement and their relationship with drift still needs to be fully understood. \citet{Engbert2006} suggested that microsaccades are triggered by a reduction in drift movement, i.e., low retinal image slip. This idea was further explored by suggesting a relationship between the self-avoiding random walk model and microsaccade triggers \citep{Engbert2011}. Our study provides further evidence supporting this connection. A build-up of activation in the SAW model's current position is associated with microsaccades. This finding is consistent with previous work indicating that a decrease in movement precedes microsaccades \citep{Engbert2006}. Our data-driven analysis revealed that the activation rises more before a microsaccade and drops more steeply compared to the randomized controls, indicating that high activation levels are more likely to trigger microsaccades. By comparing model variations, we find that this trend is primarily related to the potential, indicating that the portion of microsaccades we capture with our analysis is related to fixation control. However, further investigation is required to determine the exact mechanism behind this relationship and the potential causal direction between activation and microsaccades.

\subsection*{Other trajectory models}
Physiological drift, as the slow component of fixational eye movement, is often modeled as a random walk \citep{Burak2010, Kuang2012}. Particularly when it is considered mainly as a component in a model of visual processing \citep[e.g.,][]{Schmittwilken2022}, this approximation can yield good results. However, the statistical properties of the trajectories do differ significantly from simple randomness. To better capture these aspects, a self-avoiding random walk has been proposed \citep{Engbert2011, Roberts2013}. However, the number of models that aim to predict fixational movement trajectories is limited. The SAW model used in this paper is one example. Another self-avoiding random walk model was published by \citep{Roberts2013}. Instead of an elliptical activation trace \citet{Roberts2013} implement the self-avoidance by choosing each step direction from a continuous distribution weighted by the density of recent gaze history in each direction. It achieves a similar result: at short time scales, the model is persistent and avoids previously visited areas. The process's memory is limited and parametrized, allowing the authors insight into the process memory by estimating parameters. Due to the lack of a constraining potential, this model does not represent the subdiffusive component at long time scales. 

\subsection*{Input dependence}
Although initially fixational drift was often characterized as noise produced by the oculomotor units \citep{Eizenman1985}, more recent evidence from electrophysiological recordings in monkeys shows that fixational drift originates higher up in the chain of command than the oculomotor neurons \citep{BenShushan2022} and is influenced by attentional processes \citep{Shelchkova2019}. Microsaccades, too, are influenced by attention and preferentially move in the direction of the attended region when there is covert attention \citep{Hafed2002, Engbert2003}. Thus, drift and microsaccades depend on the viewing task and the features of the fixated target \citep{Bowers2021}. This interplay of perception and action is consistent with active vision \citep{Findlay2003}, even at the scale of fixational eye movement. 

The SAW model is stimulus-independent. Other modeling approaches have investigated the interdependence of fixational eye movements and visual perception. It can be shown that the visual processing stream is quite capable of dealing with the hypothesized motion blur caused by the constant displacement of the stimulus over the receptors \citep{Packer1992}. Fixational drift is beneficial for high-acuity vision, presumably because it allows spatial information to be redistributed into the temporal domain, modulating the input to individual receptors \citep{Clarke2021}. Image-computable models of edge detection can actually be improved by introducing drift \citep{Schmittwilken2022}. 

Research investigating the role of motion in visual perception typically uses a random walk and is most likely robust to changes in the precise type of motion. However, the experimentally observed statistical properties of drift differ from simple random noise. More research is needed to ascertain whether these properties convey additional benefits to visual processing. A recent paper \citet{Anderson2020} suggested a joint approach to infer movement and stimulus simultaneously. The model assumes a grid of retinal cells onto which stimulus patterns are projected and that the visual processing system cannot access an efference copy of the movement. Instead, they use Bayesian inference to estimate the movement and stimulus from the spike rate generated by the retinal cells alternatingly. The authors conclude that drift is beneficial for high-acuity vision as it helps the system to average over inhomogeneities in the retinal receptors. 

Thus, fixational eye movements play an essential role in visual processing. Integrating the fact that it is both stimulus-dependent and individually characteristic suggests that the movement is optimized to account for individual physiological differences. This finding is consistent with the finding that fixational eye movement and visual acuity are related \citep{Clarke2021}. A future direction for fixational drift research may be to implement the idea that drift improves visual acuity in a generative model to infer the ideal motion, prevent fading, or enable edge detection. Furthermore, although visual processing has been found to be quite robust, the development of more accurate models of fixational eye movement may improve the quality of models of visual processing.

\subsection*{Conclusion}
In conclusion, our study investigated a dynamical model of fixational eye movements and microsaccades using likelihood-based (Bayesian) parameter inference. Our analyses suggest that self-avoiding random walk models can effectively capture individual fixational drift behavior, as evidenced by the convergence of distinct marginal posteriors for each observer. Furthermore, our data-driven analysis indicates a relationship between microsaccade rate and peaks in the model's latent activation state, providing further insight into the underlying neurophysiological mechanisms of microsaccade triggering. Overall, our results provide new insights into fixational eye movements and highlight the importance of individual differences in this remarkable behavior.

\section*{Methods}
\label{sec:methods}
\subsection*{The likelihood-based modelling framework}
Biologically motivated, mechanistic models allow researchers to test whether the proposed mechanisms are capable of producing the observed behavior, to identify which components are essential, and to explore how changes to the system's structure alter its output \citep{Bechtel2010}.
Historically, the standard approach for cognitive models involves comparing them to time-independent summary statistics, e.g., here it may be MSD. 
The likelihood-based approach offers a number of advantages. First, it is possible to estimate the model parameters from the data in a fully Bayesian and statistically rigorous way. The value of the model is independent of any particular ad-hoc metric, the researcher may want to investigate \citep{Schuett2017}. The model likelihood can further be used as a basis for model comparison. Lastly, using the estimated parameters to simulate data, it is possible to conduct posterior predictive checks using metrics such as MSD to investigate whether the data constrains the model in a way that produces the expected behavior \citep{Engbert2021}. Thus, likelihood-based parameter inference allows compelling conclusions about the underlying mechanisms. Another advantage is that Bayesian parameter estimation provides a natural way to quantify uncertainty in the estimates, through the use of posterior distributions and credible intervals. This can be especially useful in cases where the data are noisy or the model is complex.

By independently estimating separate parameters for each experimental subject, it is possible to investigate individual differences. The parameter estimation yields a separate posterior distribution for each subject. As the parameters represent interpretable quantities with biological counterparts, the comparisons of the posteriors can allow interesting insights. Additionally, when a model is capable of representing individual differences, it speaks to the validity of the model and its parametrization.

\subsection*{Experimental data}
The experimental data used for this study were eye movement trajectories recorded using an Eyelink 2 with a sampling rate of 500~Hz. Participants were seated at a distance of 50~cm to the monitor and calibrated using a 9-point calibration grid. Each trial consisted of a fixation task, where participants fixated a cross in the center of a white screen for 3 seconds, followed by a scene viewing task. Here, we use only the enforced fixation data from the first 3 seconds. 
Out of 50 recruited participants, 48 completed all 40 trials. A further 6 were later excluded due to a large number of blinks. As the experiment included a  rigorous online quality control and the calibrate-ability of subjects varies, 2 participants aborted the experiment. Trials during which saccades or blinks were detected were repeated immediately. In order to detect microsaccades we used a velocity-based algorithm \citep{Engbert2015b}. The data set is publicly available on the Open Science Framework (\url{www.osf.org/hf2dw})

\subsection*{Parameter estimation}
Here, we used the DREAM\textsubscript{ZS} algorithm \citep{Laloy2012}. Rooted in the classical Metropolis (Hastings) Markov Chain Monte Carlo (MCMC) algorithm, DREAM\textsubscript{ZS} iteratively explores the parameter space by sampling its position and asymptotically converges to the true posterior distribution of the parameters. The algorithm includes several (Markov) chains starting at arbitrary (random) positions in the parameter space. For each chain new positions are chosen by combining (hence “evolution”) positions of other randomly chosen chains, including their past states. 

In Bayesian parameter estimation, the unknown parameters are treated as random variables and are assigned a prior probability distribution. This prior distribution reflects the researcher's initial beliefs about the likely values of the parameters based on prior knowledge or experience. We chose relatively broad truncated Gaussian priors, which did not constrain the estimation very strongly. The truncated tails were chosen according to experience with the model to prevent numerical problems in the case of extreme parameter values.

We split this data set into two separate sets: one half (20 subjects) was used for model development and exploratory analyses. The other half (21 subjects) was used for the final analyses and model evaluation. Each set contains data from 27 trials for each subject. We discarded all trials where the movement during fixation exceeded 1.2 degrees of visual angle. Due to the high individual variability in the data this criterion excluded 7 subjects, because too many trials were affected. The 27 trials can be split into training and test sets, with a 2/3 to 1/3 split. Each trial was 1500 samples long, i.e. represented a fixation of 3 seconds. This procedure finally yielded a data set with equal numbers of samples per trial and trials per subject, facilitating statistical analyses.

\section*{Note on comparisons of turning-angle distributions}
In order to compare and correlate the distributions turning angles of individual participants as well as between simulated and observed data, it was necessary to reduce the angle distribution to a single summary value. The area under the curve (AUC) metric is a common solution to this problem. However, in our case the distributions are densities, i.e., their AUC was equal to one by definition. Instead, we compare the AUC of the empirical cumulative density function (ECDF). The ECDF is obtained by sorting the observations into unique bins and calculating the cumulative probability for each. By grouping and computing the ECDF AUC of each group, we can compare the similarity in the peak height of the angle distributions. 
 
\section*{Acknowledgments}
This work was supported by Deutsche Forschungsgemeinschaft (DFG) via Collaborative Research Center SFB 1294, Project-ID 318763901. We thank Andrea Miroslava Pomar Robles and Stefan Seelig, who collected the experimental data set which was used in this study. 


\bibliography{main.bib}

\appendix
\renewcommand{\thefigure}{\Alph{section}\arabic{figure}}
\renewcommand{\thetable}{\Alph{section}\arabic{figure}}
\setcounter{figure}{0}
\setcounter{table}{0}

\section{Appendix}
\subsection{Discretization}
The model is defined on a $100\times 100$ lattice. In order to evaluate experimental eye movement traces (which are typically given in degrees of visual angle), we discretized the data. Each data point was multiplied by 350 and then applying the floor function. This discretization value for the entire data set was chosen by visual inspection, as it allowed all eye movement traces to stay within the confines of the grid, but also efficiently used the space. Parameter values from the stepping distribution to the potential critically depend on the specifics of this discretization.

\subsection{Simulated data}
\begin{figure}
    \includegraphics[width=\linewidth]{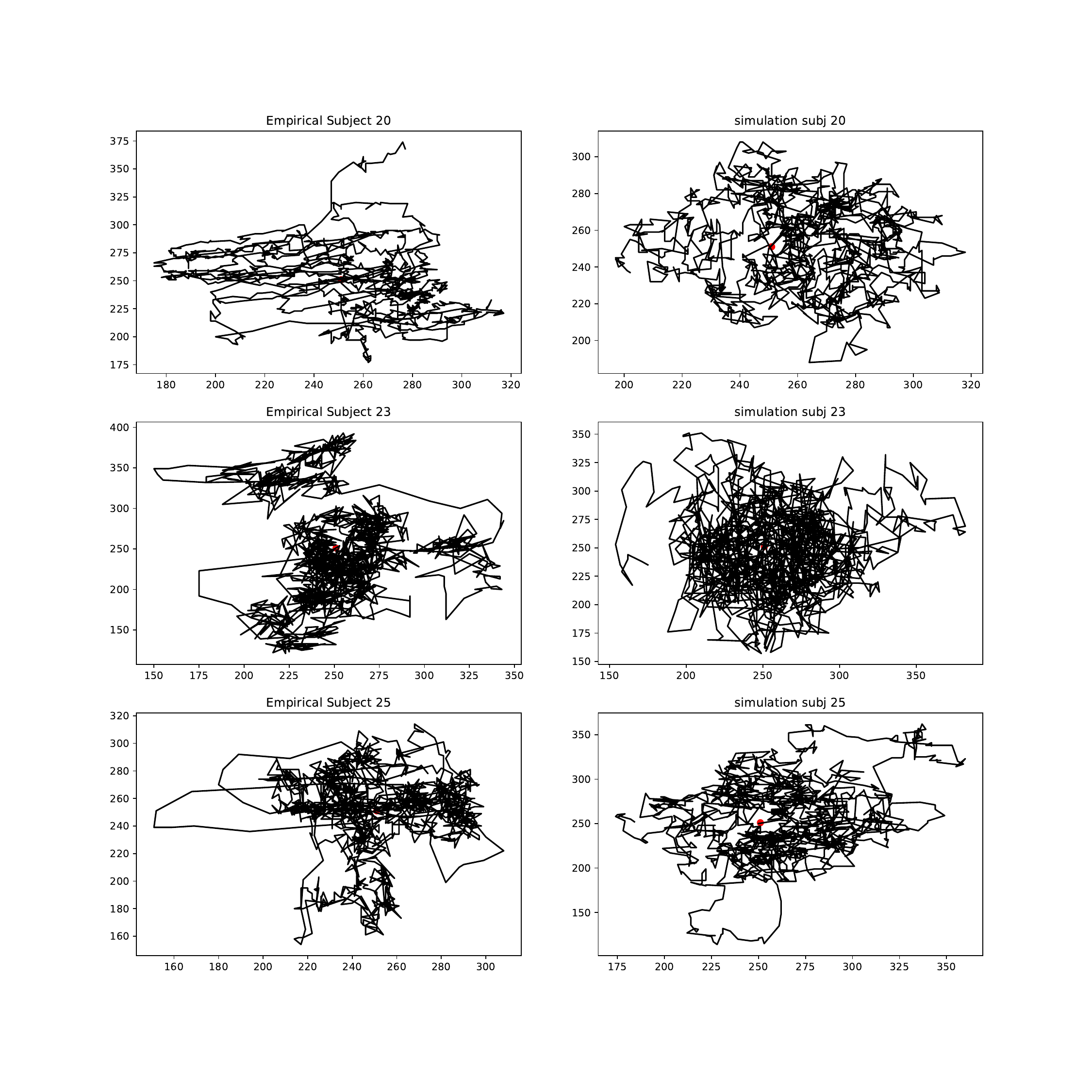}
    \caption{\label{fig:traces} Some examples of fixational eye movement traces. The left column shows the experimental data. The right column shows data simulated using the individually fitted models. These examples  qualitatively illustrate the captured statistical properties of the drift movements.
    }
\end{figure}

Figure \ref{fig:traces} shows some examples of experimental gaze trajectories which illustrate that the model captures individual differences in the data that can be validated by visual inspection.

\subsection{Parameter recovery}
Parameter recovery analyses are an important step in evaluating the stability and reliability of a mathematical model. This is often done as a first step in model building to ensure that the model is able to accurately capture the underlying dynamics of the system being studied. Parameter recovery analyses involves generating synthetic data with known (``true'') parameter values of the model and then estimating the parameters from this simulated data. This procedure permits the evaluation of the accuracy and precision of the parameter estimates, as well as the sensitivity of the estimates to the choice of estimation method and the quality and amount of data. It is an important step to assess the overall performance of the model and identify potential issues that may arise in the estimation process.

\begin{figure}
    \includegraphics[width=\linewidth]{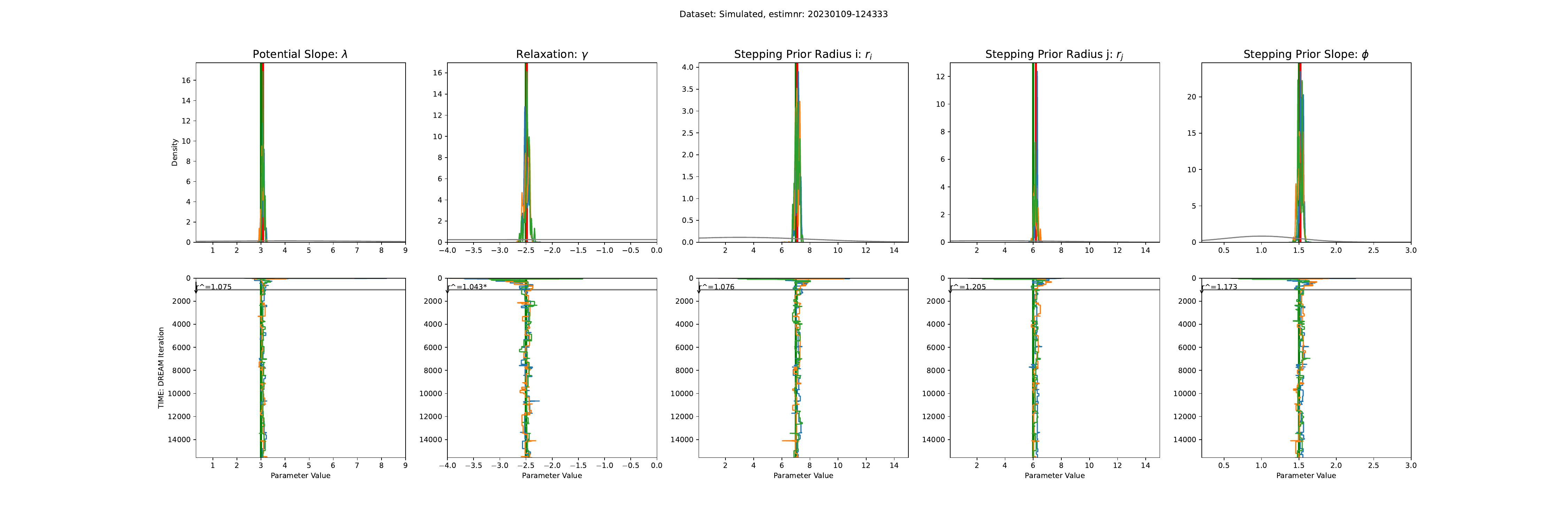}
    \caption{\label{fig:recovery}
        Parameter Recovery Analysis. For each parameter we show that a recovery analysis converges to the correct value. The top row shows the posteriors relative to the priors and reveals strong convergence in all parameters. The bottom row shows the caterpillar plots of the estimation procedure, i.e., the parameter value that the three chains assumed in each iteration. The horizontal line indicated the burn in period.
    }
\end{figure}

\subsection{Priors}
For the parameter estimation we chose truncated Gaussian priors.  Table \ref{tab:priors} details the numerical values that define the truncated Gaussian priors.

\begin{table}[h]
\centering
\small
\input{tabA1.txt}
\normalsize
\caption{\label{tab:priors} The parameters that define the prior distributions used during parameter inference. For each parameter we report the mean, standard deviation, as well as the upper and lower bounds of the truncated Gaussians.}
\end{table}

\subsection{Parameter estimation results}
Table \ref{tab:params} gives the detailed point estimates for all estimated parameters  for all participants, as well as 98\% confidence intervals. Note that the participant IDs start at 20, since we estimated parameters for the final model for participant IDs 20 to 39 only. The data for participant IDs 1 to 19 were used for model building and are omitted here to prevent overfitting.

\begin{table}[h]
\centering
\small
\input{tabA2.txt}
\normalsize
\caption{\label{tab:params} Point estimates and 98\% confidence intervals of the 5 estimated parameters for each subject.}
\end{table}

\end{document}

%% file: tabA1.txt
\begin{tabular}{x{2cm} x{2cm} x{2cm} x{2cm} x{2cm} x{2cm}}
\hline
&Potential Slope: $\lambda$ & Relaxation: $\gamma$ & Stepping Prior Radius i: $r_i$ & Stepping Prior Radius j: $r_j$ &
Stepping Prior Slope: $\phi$ \\
\hline
mean &  2   &   -2.5&    5   &   5    &   1.5    \\
sd &   2    &   0.3 &    5   &   5    &   0.5    \\
lower &0.3  &  -3.8 &    0.1 &   0.1  &   0.5    \\
upper & 8   &    0  &    15  &   15   &   3   \\

\hline
\end{tabular}

%% file: tabA2.txt
\begin{tabular}{cx{1cm}cx{1cm}cx{1cm}cx{1cm}cx{1cm}cx{1cm}cx{1cm}cx{1cm}cx{1cm}cx{1cm}cx{1cm}}
\hline
Subject&
\multicolumn{2}{x{2.5cm}}{Potential Slope: $\lambda$} &
\multicolumn{2}{x{2.5cm}}{Relaxation: $\gamma$} & 
\multicolumn{2}{x{2.5cm}}{Stepping Prior Radius i: $r_i$} &
\multicolumn{2}{x{2.5cm}}{Stepping Prior Radius j: $r_j$} &
\multicolumn{2}{x{2.5cm}}{Stepping Prior Slope: $\phi$} \\
 & mean & +/- & mean & +/- & mean & +/- & mean & +/- & mean & +/- \\
\hline
20 & 4.969 & 0.194 & -3.585 & 0.209 & 3.304 & 0.124 & 2.664 & 0.083 & 1.080 & 0.021 \\
21 & 4.497 & 0.162 & -3.688 & 0.110 & 3.850 & 0.204 & 3.378 & 0.166 & 1.072 & 0.032 \\
22 & 4.917 & 0.185 & -3.727 & 0.072 & 3.153 & 0.106 & 2.333 & 0.081 & 1.062 & 0.028 \\
23 & 4.533 & 0.045 & -3.773 & 0.027 & 12.064 & 0.572 & 8.181 & 0.376 & 1.203 & 0.055 \\
24 & 4.796 & 0.179 & -3.693 & 0.104 & 3.402 & 0.152 & 2.644 & 0.091 & 1.077 & 0.030 \\
25 & 4.563 & 0.167 & -3.757 & 0.043 & 5.156 & 0.254 & 4.172 & 0.215 & 1.010 & 0.032 \\
26 & 4.646 & 0.144 & -3.767 & 0.033 & 7.301 & 0.266 & 4.562 & 0.195 & 1.199 & 0.026 \\
27 & 5.100 & 0.163 & -3.692 & 0.105 & 2.766 & 0.143 & 2.171 & 0.109 & 0.963 & 0.031 \\
28 & 4.940 & 0.165 & -3.700 & 0.098 & 3.558 & 0.171 & 2.152 & 0.071 & 1.052 & 0.035 \\
29 & 5.211 & 0.165 & -3.723 & 0.076 & 3.353 & 0.093 & 2.700 & 0.077 & 1.111 & 0.022 \\
30 & 4.778 & 0.217 & -3.750 & 0.050 & 3.700 & 0.133 & 2.193 & 0.104 & 0.972 & 0.028 \\
31 & 4.621 & 0.115 & -3.771 & 0.028 & 7.844 & 0.296 & 5.613 & 0.209 & 1.191 & 0.031 \\
32 & 4.504 & 0.146 & -3.759 & 0.041 & 11.304 & 0.393 & 6.701 & 0.223 & 1.217 & 0.039 \\
33 & 4.565 & 0.223 & -3.696 & 0.103 & 3.527 & 0.157 & 2.499 & 0.102 & 1.012 & 0.028 \\
34 & 4.922 & 0.160 & -3.720 & 0.080 & 4.331 & 0.207 & 3.204 & 0.165 & 1.085 & 0.037 \\
35 & 4.693 & 0.209 & -3.736 & 0.061 & 3.418 & 0.125 & 2.502 & 0.087 & 1.005 & 0.021 \\
36 & 4.708 & 0.132 & -3.765 & 0.035 & 7.780 & 0.336 & 5.351 & 0.220 & 1.220 & 0.037 \\
37 & 4.223 & 0.104 & -3.755 & 0.044 & 8.409 & 0.286 & 6.516 & 0.222 & 1.211 & 0.030 \\
38 & 4.868 & 0.142 & -3.732 & 0.067 & 3.914 & 0.116 & 2.751 & 0.091 & 1.092 & 0.025 \\
39 & 5.061 & 0.091 & -3.712 & 0.084 & 4.460 & 0.188 & 3.176 & 0.128 & 1.163 & 0.035 \\
\hline
\end{tabular}